\documentclass[pre,twocolumn,groupedaddress,showpacs,showkeys,amsmath]{revtex4}

\usepackage{graphicx}
\usepackage{dcolumn}
\usepackage{bm}

\begin{document}


\title{Biased and greedy random walks on two-dimensional
lattices with quenched 
randomness: the ``greedy'' ant within a disordered environment}

\author{T. L. Mitran$^{1,2}$}
\email{tudor@solid.fizica.unibuc.ro}
\author{O. Melchert$^1$}
\email{oliver.melchert@uni-oldenburg.de}
\author{A. K. Hartmann$^1$}
\email{alexander.hartmann@uni-oldenburg.de}
\affiliation{
$^1$ Institut f\"ur Physik, Universit\"at Oldenburg, 
Carl-von-Ossietzky Strasse, 26111 Oldenburg, Germany\\
$^2$ Faculty of Physics, University of Bucharest, 077125 Magurele-Ilfov, 
PO Box MG-11, Romania
}

\date{\today}


\begin{abstract}

The principle characteristics of biased greedy random walks (BGRWs) on 
two-dimensional lattices 
with real-valued quenched disorder on the lattice edges are 
studied. Here, the disorder allows for negative edge-weights.
In previous studies, considering the negative-weight percolation (NWP) problem,
this was shown to change the universality class of the existing, static percolation
transition. In the presented study,
four different types of 
BGRWs and an algorithm based on the ant colony optimization (ACO) heuristic were considered.
Regarding the BGRWs, the precise configurations of the lattice walks constructed 
during the numerical simulations were influenced by two parameters: 
a disorder parameter $\rho$ that controls the amount of negative edge weights 
on the lattice and a bias strength $B$ that governs the drift of the walkers 
along a certain lattice direction.
Here, the pivotal observable is the probability that, after termination, a lattice walk exhibits
a total negative weight, which is here considered as percolating.
The behavior of this observable as function of $\rho$ for different
bias strengths $B$ is put under scrutiny. 
Upon tuning $\rho$, the probability to find such a feasible lattice walk increases from zero to one. 
This is the key feature of the percolation transition
in the NWP model. 
Here, we address the question how well the transition point 
$\rho_{\rm c}$, resulting from numerically exact 
and ``static'' simulations in terms of the NWP model
can be resolved using simple dynamic algorithms that have only 
local information available, one of the basic questions
in the physics of glassy systems.
\end{abstract}


\maketitle

\section{Introduction}\label{sect:introduction} 

The purely stochastic motion displayed by a simple random walk (RW)
and its various extensions and modifications that, e.g., go under the 
popular synonyms ``the ant in a labyrinth'' \cite{DeGennes1971,straley1980} and ``the biased ant in a labyrinth'' \cite{Stauffer1999},
give rise to intriguing effects such as (anomalous) diffusion, drift and 
trapping. These, and further related topics, were actively studied since 
the beginning of the last century, first by analytic means and later using
extensive numerical simulations. A subset of those studies, relevant in the
context of the presented article, is reviewed in subsect.\ \ref{sect:literatureReview}.
In all of these studies, a random walker is considered that traverses 
a given (disordered) environment subject to different dynamics-governing rules.

Here, we also consider a random walker on a disordered 
``energy'' landscape, where, 
in contrast to most previous studies, the underlying energy-values 
might have a positive
or a negative sign. Note that the existence of the negative
energies fundamentally changes the behavior of the model defined below and
will be destroyed by lifting all energies by the same amount to a positive
value. The reason for this fundamental difference 
is that, in the presence of a suitable amount of 
(non-replenishable) negative
energies, the walker might lower its energy more and more by performing longer
and longer walks, in contrast to the presence of only positive energies.
The energy values are drawn from 
specified distributions 
wherein a disorder-parameter $\rho$ allows to alter the fraction 
of negative energies in the
environment (see discussion below).

Regarding such a disordered environment, we have previously introduced 
\cite{melchert2008} and investigated \cite{apolo2009,melchert2010a,melchert2011b,Norrenbrock2012,claussen2012}
the \emph{negative-weight percolation} (NWP) problem by numeric means.
Regardless of the spacial dimension of the underlying (hypercubic) lattice
graph, the observables in the NWP problem are always line-like, i.e.\ have an intrinsic dimension 
of $d=1$. 
As observables in the NWP problem one considers, e.g., minimum-weight paths 
(in the presence of a possibly empty set of negative weighted loops) that
span the underlying lattice between a pair of specified boundaries.
The problem of finding these paths  numerically
can be cast into a minimum-weight
perfect matching problem, e.g.\
discussed in Refs.\ \cite{melchert2010a,Norrenbrock2012} in more detail (in the latter reference, 
a comprehensive description and illustration of the respective algorithm can be found).
A pivotal observation is that, as a function of the disorder parameter 
$\rho$,  
the NWP model features a disorder-driven phase transition, at which the actual minimum-weight path energy becomes
negative. In this question, note that the aforementioned transition in the path energy is a precursor
to a further ``geometric'' phase transition at which the paths exhibit a roughness of the 
order of the systems size and where loops might appear that span the system along at least one
direction \cite{melchert2008}.
In the limit of large system sizes, there is a particular 
value of the disorder parameter, signified as $\rho_c$, at which paths with a negative energy appear 
for the first time. E.g.\, for the  square 
($2D$) lattice, considering a bimodal disorder distribution, 
the average path energy turns negative above $\rho_c=0.0869(2)$ \cite{melchert2008}.
Therein, the finite-size scaling behavior is described by the critical exponent $\nu=1.47(6)$,
consistent with the one that describes the geometric transition of the loops-only setup.
At $\rho=0.1032(5)$, slightly above this critical threshold, the fractal dimension of the paths was estimated to be $d_{\rm f}=1.268(1)$.
Further, at this critical point, the loops that one finds in addition to the minimum-weight
path are rather small and isolated.
Hence, they (presumably) do not affect the precise configuration of the minimum-weight paths.

As mentioned above, NWP was previously studied numerically exact by employing a mapping to an auxiliary 
minimum-weighted perfect matching problem, which relies on an involved optimization algorithm specifically tailored 
for the problem at hand. 
Instead of using such a complex algorithmic approach that finds the optimum of a given optimization
problem using \emph{global} information, we here address the question how well the aforementioned transition point $\rho_c$ (resulting from numerically exact simulations 
that involve a high degree of optimization) can be resolved using simple dynamic algorithms that have only 
\emph{local} information about the assignment of edge weights available. 
This is a more physical viewpoint on the problem,
where, typically, the movement of particles is mostly determined by local
information. Hence, a comparison between global static and dynamical 
physical viewpoints is possible. These two perspectives are
often taken for the study of disordered and glassy systems 
\cite{young1998,phase-transitions2005,binder2011}.
Therefore we consider different biased random walk approaches and study
their ability to find system spanning paths with negative path-weight.
Averaging over different realizations of the disorder, we compute the probability to find a negative weighted path as a function 
of the model parameter $\rho$ for different simple dynamics.
The algorithms we consider here all mimic moving particles (also referred to as agents or walkers) in a disordered environment. The simplest dynamics we
study stems from a biased random walk on the lattice and the most intricate dynamics is implemented in terms
of a particular heuristic algorithm known as ``ant colony optimization''.
Subsequently, we will refer to $\rho_c$ as the ``dynamic'' transition points, as opposed to the 
result from the ``static'' (global) simulations quoted above. 
Not too earnest one might allude to this study as being dedicated to the behavior of
``the greedy ant in a disordered environment''.

The remainder of the presented article is organized as follows.
In section \ref{sect:literatureReview} we review some of the 
related literature that alludes to effects that we also expect to observe within
our simulations. In section \ref{sect:model}, we introduce the model in 
more detail and we outline the different dynamic algorithms used to 
perform the lattice walks. In section \ref{sect:results}, we list the results of 
our numerical simulations and in section \ref{sect:conclusions} we 
conclude with a summary.

\section{Review of related literature}\label{sect:literatureReview}
A typical observable for the characterization of ordinary RWs is given by
their mean square displacement (MSD) $\langle R(t)^2\rangle$, which, as a function
of time $t$, is expected to scale as
$\langle R(t)^2\rangle = D t$. Therein, $D$ signifies the diffusion 
coefficient characteristic for the observed walks.

A particular extension of the simple RW model, relevant in the 
context of the presented study, consists in the addition of disorder
to the environment in which a walk is performed. Different approaches
to implement disorder can be found in the literature: 
E.g., one might consider percolation-like ``occupied vs.\ free'' site (or 
bond) disorder \cite{benavraham1982,gefen1983,padey1984,argyrakis1984No2,white1984,havlin1987}, in which 
case a fraction $p$ of lattice sites (or bonds) is distinguished as ``occupied'', and where a RW
is performed on a cluster of occupied nearest neighbor sites 
(a.k.a.\ ``the ant in the labyrinth''). For short walking times, 
i.e.\ times within which the walker cannot fully trace the cluster, and 
in the vicinity of the critical point $p_c$, signaling the onset of percolation, 
\emph{anomalous diffusion} was observed \cite{gefen1983,bouchaud1990}.
Thereby, the MSD of the RW scales according to $\langle R(t)^2\rangle \propto t^{\alpha}$, 
characterized by an effective dimension $d_{\rm RW}=2/\alpha$. At the critical point 
for site dilution on $2D$ lattices, Ref.\ \cite{benavraham1982} reports $d_{\rm RW}=2.68(5)$, i.e.\ $\alpha\approx 0.75$
indicating a sub-diffusive behavior. 
For the non disordered case (i.e.\ in the limit $p=1$) one would expect to find $d_{\rm RW}=2$, 
characterizing the effective ``fractal'' scaling dimension of unhindered RWs.

A different approach considers ``energetic'' disorder associated with sites or bonds \cite{argyrakis1984,avramov1993},
representing random barriers that define site-to-site transition 
probabilities governed by Boltzmann statistics. Therein, a model inherent temperature-parameter
$T$ controls the ability of an individual walker to overcome energy barriers. 
At low temperatures, a RW is likely to get trapped at local minima in the 
energy landscape. Albeit this trapping effect is only temporary, the RWs generally
display a rather limited mobility. With increasing temperature the RWs gain mobility 
and behave similar to ordinary RWs in the limit $T\to\infty$.
Regarding this energetic disorder
one might further distinguish static (i.e.\ quenched) disorder, which remains 
unchanged in time, and dynamic 
disorder, which is renewed after each discrete time step, see Ref.\ \cite{avramov1993}.
In case of static disorder and after a temperature dependent \emph{crossover timescale} $\tau_c$, the
MSD exhibits the same asymptotic linear scaling as for unhindered RWs.
Supported by analytical arguments from classical chain-reaction theory, the crossover time
$\tau_c$ could be attributed to an effective activation energy that
must be overcome in order to escape from an initial local minimum in the energy landscape.
Intuitively, the lower the temperature, the larger the crossover time tends to be. In the early time
regime, i.e. at times smaller than $\tau_c$, the scaling of the mean square displacement
is sub-diffusive.
In case of dynamic disorder, no crossover time is found and the same scaling as for the 
unhindered RW is observed for all temperatures. The renewal of the disorder after each 
time step creates an averaged environment in which the temperature $T$ directly reflects 
the transport efficiency in the medium, resulting only in a temperature dependent diffusion constant $D(T)$.

A further extension that is heavily studied in the literature consists in emphasizing a particular lattice
direction along which the RW effectively drifts. Also
for this modification there are different
possible implementations, differing in the particular way in which the transition probabilities
are altered, see \cite{bunde1987,arapaki1997,avramov1998,dhar1998,Stauffer1999,kirsch1999,zhang2006}.
In this regard we only briefly recap one of those studies, which, for the first time, reported on a
direct observation of a sharp \emph{drift to no-drift} transition \cite{dhar1998}.
Therein the influence of a directional bias field on the diffusive behavior of random walks on $3D$ lattice graphs with
percolation-like site disorder at $p=0.5$, i.e.\ above the critical point $p_c\approx0.3116$ of the $3D$ setup, was addressed. 
Initially, for a given realization of the disorder, noninteracting RWs were started on randomly chosen occupied sites 
and individual walkers were only allowed to advance to adjacent occupied sites. 
A bias parameter $0\leq B\leq1 $ is used to control the drift of the RWs along the x-direction: 
the probability to advance along the positive x-direction is set to $B$, the probability to advance 
to any of the six neighbors is set to $1-B$. Once such a step is proposed, it is only executed if the target site
also belongs to the cluster. Hence, there is the possibility that a walker remains at its current position 
(speaking in terms of the popular phrases mentioned in the introduction, it would be more precise
to call this the ``blind biased ant in the labyrinth'' since the decision on the direction of the proposed
move is made irrespective of whether the target-site can actually be reached).
For such noninteracting walkers and prior to the numerical study reported in \cite{dhar1998}, 
a drift to no-drift transition at a finite critical value $B_{\rm c}$ was 
theoretically hypothesized \cite{barma1983,dhar1984}.
For weak bias $B$, the walkers exhibit a diffusive motion and their velocity $v(t)$ 
assumes a constant value which tends to increase with $B$.
For strong bias, diffusion is slowed down since the walkers tend to wind up in ``dangling ends''
of the clusters on which they reside. With increasing bias they have an increasingly hard
time to escape from such dangling ends. 
Neglecting initial transients, three different regimes were observed:
(i) $B<B_{\rm c}$, where $v(t)$ tends to a constant,
(ii) $B=B_{\rm c}$, where $v(t) \sim 1/\log(t)$, and,
(iii) $B>B_{\rm c}$, where $v(t) \sim t^{-x(B)}$ (and $x(B)>0$). The critical bias strength at which the
drift to no-drift transition occurs was estimated as $B_{\rm c}\approx 0.53$.
The behavior of the average velocity for a large range of $B$-values, covering all three
scaling regimes, could further be summarized by an extrapolated scaling form.
Similar studies were, e.g., also carried out to construct a full phase diagram for the 
drift to no-drift transition in the $B$-$p$--plane for the $2D$ square lattice \cite{kirsch1999}.
Regarding the MSD, the RW component of the dynamics leads to diffusion $\langle R(t)^2 \rangle\propto t$
and the bias induces a drift $\langle R(t)^2 \rangle \propto t^2$. At this point note that under
drift, the asymptotic effective dimension $d_{\rm RW}=1$ might be expected. At short (long) times diffusion (drift) is dominant.
Under a strong bias, the dangling ends act as traps that inhibit drift by characteristic
``trapping times'' (depending on the precise cluster structure). Consequently, the MSD is
governed by effective exponents $\alpha_{\rm eff}(p)$ that approach their asymptotic value 
rather slowly \cite{pandey1984,dhar1984,kirsch1999,Stauffer1999}.
Further, biased diffusion for networks without underlying regular spatial
structure (i.e. regular random graphs, Erd\H{o}s-Renyi random graphs and
scale-free networks) was recently investigated by means of numerical
simulations and analytic calculations \cite{Skarpalezos2013}. Therein, the bias
was designed in a way that it guides the walker to a specified target node
along the respective shortest path. Using a model parameter, the strength of
the bias, ranging from unbiased (yielding an ordinary random walk dynamics) to
fully biased (effectively confining the walker to the shortest path), could be
tuned.  Among other things it was shown that for regular random graphs and
Erd\H{o}s-Renyi random graphs, the scaling of the mean first passage time as
function of the system size changes from a power-law to a logarithmic behavior
at a characteristic value of the bias parameter.

In addition to simple RWs, self-avoiding walks (SAWs), conveniently used as models 
for randomly bent polymers that exhibit an ``excluded volume'' effect \cite{caracciolo1992}, have been studied for
diluted systems exhibiting percolation
disorder \cite{kremer1981,grassberger1993}. Therein, a basic observable is the 
mean square radius $\langle R^2 \rangle$ of $N$-step SAWs which scales according to $R_N\propto N^{2 \nu_{\rm SAW}}$, 
where the exponent $\nu_{\rm SAW}$ can be seen as an inverse of a SAW characteristic scaling exponent $d_{\rm SAW}=\nu_{\rm SAW}^{-1}$
that is universal in each dimension $d$. Depending on the fraction $p$ of occupied sites 
on the underlying lattice and the set of configurations over which averages are computed, different SAW scaling regimes can be found \cite{grassberger1993}. 
For the case of no disorder (equivalent to $p=1$), results consistent with
$\nu_{\rm SAW}=0.75$ ($d_{\rm SAW}=1.333$) are obtained
in two dimensions, see, e.g., Ref.\ \cite{beretti1985}.
A different study, aimed at clarifying the geometric and energetic scaling behavior of minimum energy (ME) SAWs on
lattices with random site energies \cite{smailer1993},
reports $\nu_{\rm ME-SAW}=0.80(2)$ ($d_{\rm ME-SAW}=1.25(3)$) for the $2D$ case (obtained via exact enumeration methods). Thus, under the influence of quenched
randomness, minimum energy SAWs tend to expand. Note that within errorbars this value is 
consistent with the fractal dimension $d_{\rm f}=1.268(1)$ of paths in the NWP problem at the 
particular point $\rho=0.1032(5)$ (also in $3D$ the estimates of the respective exponents stack up well: $d_{\rm ME-SAW}=1.41(6)$ 
as compared to $d_{\rm f}=1.459(3)$ \cite{melchert2010a}).

A further variant of a simple RW, wherein a loop is erased as soon as it is formed, is referred to 
as loop-erased random walk (LERW) \cite{majumdar1992}. A LERW might be interpreted as a simplified version of a SAW
which has some correspondence to spanning trees and, in $2D$, exhibits a scaling dimension of $d_{\rm LERW}=1.25$ \cite{grassberger2009,majumdar1992}.


\section{Model and algorithms}
\label{sect:model}

\begin{figure}[t!]
\centerline{
\includegraphics[width=1.0\linewidth]{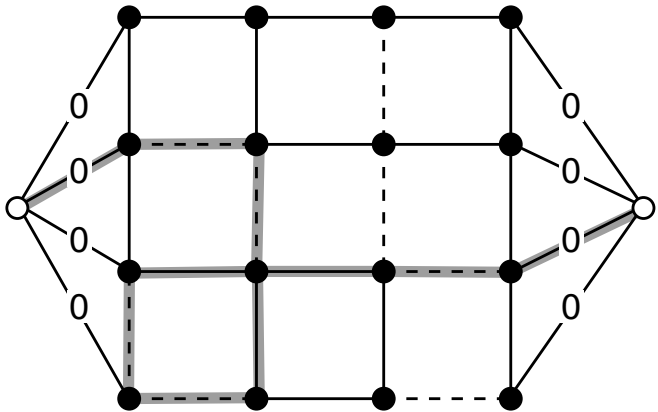}}
\caption{Example of a lattice walk on a disordered $2D$ square lattice
of side length $L=4$ and open boundaries in the vertical direction
(note that in the actual simulations, periodic boundary conditions in this
direction are considered). The solid (dashed) lines indicate edge weights
$1$ ($-1$) and the edges that connect the outer nodes (open circles) to the
nodes on their respective boundaries carry zero weight. 
The fraction of negative edge weights it thus $\rho\approx 0.29$. The bold grey 
line indicates a possible exemplary lattice walk of weight $\omega_{\rm path}=-2$.
\label{fig:latticeWalk}}
\end{figure}  

In the present article we consider lattice graphs
$G\!=\!(V,E)$ with a $2D$ square lattice geometry, 
having side length $L$ and periodic (free) boundary 
conditions (BCs) in the vertical (horizontal) direction.
The considered graphs have $N\!=\!|V| \!=\!L^2$ nodes 
(plus two extra ``outer'' nodes) 
$i\!\in\!V$ and $M\!=\!|E|\!=\! L(2L-1)$ undirected edges 
$\{i,j\}\!\in\!E$  that join adjacent nodes $i,j\!\in\!V$ 
on the lattice, see Fig.\ \ref{fig:latticeWalk}.  
In addition, there are $2L$ extra edges that attach the 
$L$ nodes on the left and right boundary to the respective outer nodes.
We further assign a weight $\omega_{ij}$ to each $\{i,j\}\in E$. 
These weights represent quenched random variables that introduce disorder 
to the lattice.
Here we consider independent identically distributed 
weights drawn from (i) the \emph{bimodal} distribution 
\begin{equation}
P_1(\omega)=(1-\rho)~\delta(\omega-1)+\rho~\delta (\omega+1), 
\label{eq:disorderBimodal} 
\end{equation} 
where $\rho$ signifies the fraction of negative edge weights $\omega=-1$, or
(ii) the \emph{semi-continuous} distribution
\begin{equation}
P_2(\omega)= (1-\rho)~\delta(\omega-1) + \rho~u(-1,0), \label{eq:disorderSemiContinuous}
\end{equation} 
wherein $u(-1,0)$ signifies a random number uniformly drawn from the interval
$(-1,0]$.

Subsequently we will study lattice walks on instances of weighted $2D$ lattice 
graphs with sidelength up to $L=800$. The considered walks are confined to start at an 
outer node attached to the left lattice boundary and end as soon as the walker reaches an
outer node attached to the right boundary, see Fig.\ \ref{fig:latticeWalk}.
More precisely, to construct the lattice walks, two classes of algorithms were used: 
\emph{biased and greedy RWs} and an \emph{ant colony optimization} (ACO) heuristic. 

\subsection{Computing paths via biased and greedy random walks (BGRWs)} \label{subsect:BGRW}

In brief, the biased RWs explore the lattice graph while greedily choosing 
the ``best'' direction at each discrete timestep. This might be the lowest 
weighted edge connected to the node the walker currently resides on, or, if 
there is a draw, one of the lowest weighted edges picked at random. 
By ``biased RW'' it is meant that following a certain rule, 
a preferred direction 
is chosen for the next step, even if it is not the best possible immediate choice.
The rule employed during our simulations reads as follows: 
say the preferred direction is given by the positive x-direction. First, the 
walker determines the direction corresponding to the best local edge-choice.
If this direction is already the positive x-direction, the walker goes there.
Otherwise, if the best local choice does not lead towards the positive x-direction,
the walker accepts the choice only with probability $1-B$, where $B\in[0,1]$ (termed ``bias probability'' or
short ``bias''). I.e., if its not the best local choice, the walker goes towards 
the positive x-direction with probability $B$.

Thus, a walker experiences a drift that, depending on the 
bias-direction, effectively ``guides'' him towards a particular direction (see discussion in 
sect.\ \ref{sect:literatureReview}). Here, we will refer to this kind of RW in 
a disordered environment as ``biased and greedy random walk'' (BGRW).
While traversing the graph, the walker keeps track of the edge weights it encounters.
Consequently, a lattice walk is considered successful if the total sum of the edges traversed
by the walker is smaller or equal to zero. Clearly, the walker will
be able to find negative-weight paths only if they exist, which can be
detected by the previously mentioned exact algorithms. We anticipate that
there will be an intermediate range, where negative-weight paths exist,
but are so rare that they cannot be detected by the locally-acting walker.

We also study a variant of such lattice walks, where the walker interacts with the graph 
by modifying the weights of the traversed edges.
This is to some extent motivated by studies, as, e.g., the one reported in Ref.\ \cite{boyer2004}, aimed
at mimicking the foraging behavior of social monkeys. 
Here, we chose the rules that the walker 
\begin{itemize}
\item[(i)] replaces both, positive and negative edge weights 
by weight $\omega=0$ (aimed at modeling a finite non-replenishable amount of 
both, resources and costs), or 
\item[(ii)] only replaces negative edge weights with the standard 
positive value of $\omega=1$ (aimed at modeling non-replenishable
finite resources, only).
\end{itemize}
These types of modifications where chosen to also prevent the walker from getting stuck in regions
of the lattice with a high density of negative edge weights, as for RWs in disordered energy 
landscapes at low temperatures reviewed in sect.\ \ref{sect:literatureReview} above.
If no such interaction with the lattice is implemented, the walker could in principle
collect unlimited amounts of negative edge weights.

Four distinct variants of BGRWs, termed RWs of type A, B, C or D, were considered to 
evaluate which one performs better in finding paths with an overall negative weight.
Here, the term ``better'' means ``with higher probability for a given disorder parameter
value $\rho$''. Also bear in mind that ultimately, the 
aim is to quantify to which extent the 
transition point $\rho_c$ (known from numerical simulations 
that involve finding the exact global optimum \cite{melchert2008}) 
can be resolved using simple dynamic algorithms 
that have only \emph{local} information about the assignment of edge weights available.
The four variant of biased RWs, along with a brief description of the 
dynamic governing rules, are listed below. The variant A is 
based on standard RWs, while the
other three variants are based on (partially) loop erased RWs.

\begin{figure}[t!]
\centerline{
\includegraphics[width=1.0\linewidth]{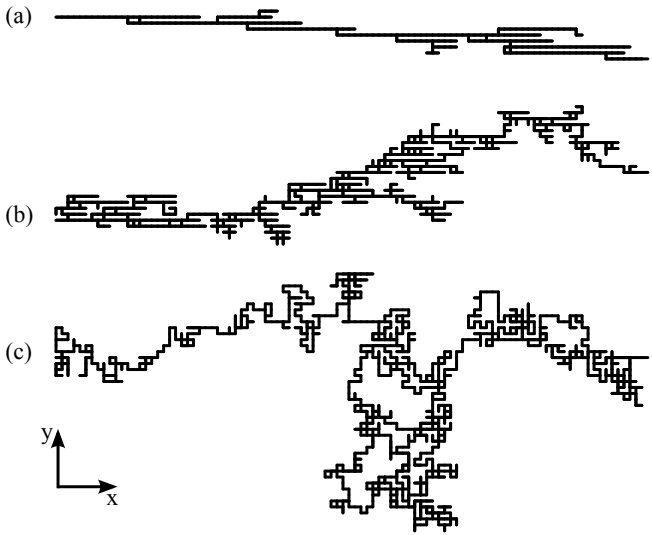}}
\caption{Examples of type {\rm A} lattice walks on a disordered $2D$ 
square lattice (considering bimodal disorder) of size $L=100$ for different values of the disorder parameter $\rho$. 
In this example, while traversing the lattice graph under a bias parameter $B=0.05$ , the walker modifies
positive and negative edge weights, replacing them by edge weights with value
zero (i.e.\ rule (i) described in the text).
The subfigures correspond to (a) $\rho=0.1$, (b) $\rho=0.5$, and (c) $\rho=0.9$.
\label{fig:latticeWalks_typeA}}
\end{figure}  

\begin{itemize}
\item {BGRWs of type {\rm A}:} this type of walk is given by a 
BGRW, that replaces 
negative edge weights (rule (ii) above) and maintains a running sum of 
the edge weights encountered while traversing the graph. 

For completeness, note  that 
also the variant wherein the walker replaces edge weights according to rule (i) was 
considered, see Fig.\ \ref{fig:latticeWalks_typeA}.
In the figure, type {\rm A} walks on a square lattice
of size $L=100$ for different values of the disorder parameter $\rho$ 
are shown,
considering the bimodal disorder distribution $P_1(\omega)$ (see Eq.\ \ref{eq:disorderBimodal})
for different fractions of negative edge weights $\rho$.


While traversing the lattice graph, the walker modifies
positive and negative edge weights, replacing them by edge weights with value
zero. As a consequence, for a small fraction of negative edge weights 
(as for $\rho=0.1$ in 
Fig.\ \ref{fig:latticeWalks_typeA}(a)) it can be observed that the walker
backtracks along subpaths of considerable length. This is due to the fact that, 
after the walker modified the respective edge weights, they are actually ``cheaper''
than the $+1$ edge weights that characterize the majority of the edges.
The bias is designed to induce a drift along the positive x-direction, hence one finds 
a pronounced ``channeling effect'' along the horizontal lattice axis.
Once such a region of zero weighted edges develops, for 
small values of $\rho$,
the walker has a hard time to escape
from it. As evident from Figs.\ \ref{fig:latticeWalks_typeA}(b,c), 
the channeling effect is less pronounced as the fraction of negative edge weights increases.
Note that this effect is conceptually similar to the 
behavior of RWs in energetically disordered lattices, reviewed in sect.\ \ref{sect:literatureReview}
(see Refs.\ \cite{argyrakis1984,avramov1993}),
where initially an effective activation energy must be overcome in order to escape from 
a local minimum in the energy landscape.
The general characteristics of such a lattice walk, basically confined to take place 
on edges of zero weight in an environment that mostly features positive edge weights, further 
bears some resemblance to the trapping behavior observed for RWs on percolation clusters
discussed previously.

\item {BGRWs of type {\rm B}:} this type of walk is given by a 
loop-erased BGRW, traversing the graph without modifying the edge weights
(which is not necessary, due to the erasure of the loops).
The sum of edge weights along the loop-erased walk is computed as soon as the BGRW has finished.

\item {BGRWs of type {\rm C}:} this type of walk combines an interaction with 
the environment (as for type {\rm A} walks) with loop-erasure (as for type {\rm B} walks).
The lattice walk is constructed via the following two-step procedure:
In the first step, the walker performs a BGRW and modifies the edge weights related 
to the traversed edges according to rule (i) above. At each visited node it further stores 
the sum of edge weights accumulated so far, the previous node it has visited, and the 
original weight (i.e.\ the value of the edge weight before it was
modified by the walker) of the edge it has just traversed. 
In the second step, after the BGRW has terminated, the loop-erasure and subsequent
path evaluation is performed. Therefore, the lattice path is traced back, starting from 
the final node, by selecting at each intermediate node the step that yields 
the lowest sum of the edge weights until that point.
This selection appears each time when a loop is encountered
while backtracking. This is a heuristic which leads to lower
path weights, but is not able take combinatorial cases like a loops
inside loops inside loops, etc., into account, since this would contrast the idea
of using fast and local heuristic algorithms for computing the walks. 
Finally, the total weight of the path is computed by summing up the edges. 
For completeness, note  that 
also the variant wherein the walker replaces edge weights according 
to rule (i) was considered.

\item{BGRWs of type {\rm D}:} this type of walk might be referred to as
``partially'' loop erased BGRW, wherein only unfavorable 
loops along the lattice
walk are erased. More precisely, the walker modifies the weights of the traversed edges and checks each time
if the node it currently resides on was visited previously. If so, the walker 
checks whether the created loop has a negative or positive weight. The loop
is kept as part of the lattice walk if it has a negative weight, 
or, if it has a positive weight, the loop is discarded and the original edge weights are restored along it. 

Regarding these dynamics, only the variant where the walker replaces edge weights
according to rule (ii), 
i.e., the modification of negative edge weights only, was considered.
This was done
since the modification of positive and negative weights led to the walker being stuck for a very 
long time in a certain part of the graph, 
especially when using a low value for the bias and a small fraction of negative weights. 
The cause for this was that the walker created areas of zero weight in the graph that it 
could not escape efficiently if the surrounding edges were all positive and the bias 
was too low (somehow similar to the effective activation energy that must be overcome
in the initial phase of RWs in energetically disordered lattices reviewed in sect.\ \ref{sect:literatureReview}). 
A similar horizontal ``channeling'' effect can be seen in Fig.\ \ref{fig:latticeWalks_typeA} for random walks of type {\rm A} BGRWs,
where the walker is constrained to move on an almost straight line because of the ``repulsive'' effect
of the positive edge weights. 
\end{itemize}

\subsection{Computing paths via ant colony optimization (ACO)}\label{subsect:ACO}

Previously we considered algorithms that took into account \emph{local} information
only. I.e.\ the \emph{random walk} approaches we considered so far made a decision on 
where to go next by only considering the weights of the edges adjacent
to the walkers current location.

In this subsection we will consider a nature inspired, population based heuristic for the solution 
of optimization problems, called \emph{Ant Colony Optimization} (ACO) \cite{ACO_homepage,Dorigo2005_aco}.
Designed to mimic the foraging behavior of actual ants, the ACO heuristic 
is a particular example of swarm intelligent systems,
which has proven to be valuable in solving a wide range of optimization problems 
that can be cast into the form of optimal-path problems. 
As regards this, the ACO heuristic has already been 
applied to different optimization
problems such as, e.g., the traveling salesperson problem 
\cite{Dorigo1997_aco}, which is a hard optimization problem, 
and the  
minimum-weight spanning tree problem \cite{Neumann2010_aco}, which
is polynomially solvable, i.e., easy.

The ant \emph{colony} represents a population of, say, $M$ individual agents
that are all able to construct solutions to the given problem by considering
local information only (these need not necessarily be solutions of high quality). 
Albeit the $M$ agents do not interact directly, they are able to interact indirectly
through the deposition of \emph{pheromone} on the graph edges (see discussion below). 
These interactions should lead to lower-energy solutions as compared
to the single random-walk algorithms discussed above.

Here, the problem is to find a loopless minimum weight path from a 
specified source node $s$ to 
a specified target node $t$ in a weighted undirected graph $G=(V,E)$, as discussed earlier. 
The ACO heuristic comprises an iterative algorithm, where in the beginning, 
there is no pheromone on the edges $\{i,j\}\in E$. Hence,
the only information associated with a particular edge is its weight $\omega_{ij}$.
A basic variant of the ACO algorithm for the above problem can be cast into the following steps: 
\begin{itemize}
\item[(i)] Preprocessing of the edge weights: Transform the set of edge-weights $\omega_{ij}$ to a set of (initial) transition probabilities 
$\tau_{i\to j}= Z_{i}^{-1} \exp\{-\omega_{ij}\}$, where $Z_{i}=\sum_{j\in {\rm Nb}(i)} \exp\{-\omega_{ij}\}$ with
${\rm Nb}(i)$ specifying the set of nodes adjacent to node $i$. In this way it is assured that
an edge weight $\omega_{ij}$ with a comparatively large negative weight will result in a comparatively 
high probability for the agent to take the step $i\to j$. These initial transition probabilities can also be 
thought of as an initial distribution of pheromone which affect the behavior of the individual agents. 

\item[(ii)] Generate solution to the problem: The ability of an individual agent to construct loopless paths as solution to the 
problem at hand can be implemented in various ways. For a more clear illustration lets
consider one agent only, i.e.\ $M=1$. Here, starting at a specified source node $s$, we 
let the agent perform a loop erased random walk (LERW), guided by the transition probabilities $\tau_{i\to j}$.
During the loop erased walk, the agent sums up the edge weights along the LERW edges.
As soon as the agent reaches a specified target node $t$, the guided LERW is completed and 
a solution to the problem, i.e.\ a loopless $(s,t)-$path, is obtained. Note that this is not necessarily a solution with a good
quality regarding the given optimization criterion. 

\item[(iii)] Evaluate quality of the solution and deposit pheromone:
In order to quantify the quality of the LERW obtained in step (ii) and so as to 
reflect the optimization criterion of the problem at hand, we compute the ``fitness'' parameter
$q=\alpha (1- \omega_{\rm p}/\ell_{\rm p})$, where $\alpha>0$ is a tunable parameter, while 
$\omega_{\rm p}$ and $\ell_{\rm p}$ are the weight and length of the LERW, respectively. 
Note that for edge weights drawn from a bimodal distribution, it holds that $q\geq 0$. 
Further, note that the more negative the sum of the edge weights, the larger the value of 
$q$ is. 
Then, modify the transition probabilities along the LERW to $\tau_{i\to j} \to \tau_{i\to j}+q$ and 
normalize them as in step (i) above. This is called a delayed pheromone update, since 
the local information on the edges is updated after the LERW is obtained. 
The parameter $\alpha$ can be tuned to alter the influence of the quality measure $q$ on the subsequent recomputation
of the transition probabilities.
Here, we optimize the ratio $\omega/\ell$, effectively measuring the average
edge-weight that contributes to the path. This quantity was chosen since it
naturally normalizes the value of $q$ to the range $q\in[0,2\cdot \alpha]$ and
thus allow to control the influence of the current local solution on the
transition probabilities in an easy way by using the parameter $\alpha$
(in contrast, note that in the static NWP problem we optimize $\omega$ instead).

\item[(iv)] Evaporate pheromone:
While pheromone is deposited only on the edges of particular loopless paths, it can be useful to also 
evaporate some of the pheromone from all edges. E.g., a simple evaporation heuristic is to 
modify the transition probabilities along all edges to $\tau_{i\to j} \to \beta\cdot \tau_{i\to j}$, where $\beta\in [0,1]$, and 
to normalize them as in step (i) above. 
\end{itemize}

Step (i) is a preprocessing step to generate transition probabilities that reflect the
weight assignment on $G$.  Steps (ii) and (iii) complete the ``lifecycle'' of an individual
agent: It generates a (non-optimal) solution to the optimization problem and evaluates
a fitness for the solution. The fitness is then used to alter the transition probabilities
which are associated with the edges. Instead of using just one agent, steps (ii) and (iii)
straightforwardly generalize to $M$ individual agents. The construction and evaluation of
$M$ LERWs and the subsequent pheromone evaporation, i.e.\ steps (ii-iv) comprise one \emph{sweep}. 
Consequently, the ACO algorithm is iterated for a number
of, say, $n$ sweeps. The transition probabilities change from sweep to sweep. 
Thereby, edges that belong to $(s,t)-$paths with a high fitness get equipped with a larger
transition probability. Albeit this induces a positive feedback that allows to distinguish 
``good'' paths, the parameter $\alpha$ can be used to limit the extent to which the transition
probabilities are modified from sweep to sweep (during our simulation we used $\alpha=1/M$). 
Also note that the parameter $\beta$ serves to limit the ``long-term memory'' of the 
search process in order to efficiently explore a large
variety of $(s,t)-$paths (during most of our simulations we used $\beta=0.98$).
E.g., setting $\alpha=0$ and $\beta=1$ completely suppresses the deposition and evaporation
of pheromone. For that parameter choice the individual agents would construct LERWs which 
are guided only by the transition probabilities induced by the edge-weights on the lattice (following step (i))
and they would not be able to indirectly communicate through the deposition of pheromone.
Also note that in the extreme case where $\alpha\to \infty$ and $\beta=1$, 
the edges $\{i,j\}$ that belong to a $(s,t)-$path 
found in step (ii) would get a transition probability $\tau_{i\to j}\approx 1$ and all transition probabilities 
that relate to those edges that depart from that particular path would effectively be zero. Hence, 
once an agent crosses that path (during a later sweep), it is immediately ``confined'' to that path and is unlikely to 
escape from it. 
During a simulation run for a particular realization of the edge weights we store the best path
found so far. This path is returned by the ACO algorithm after it terminates.


\section{Results}\label{sect:results}

As pointed out in the introduction, minimum weight paths on disordered lattices were
already studied in terms of the NWP model, which, for a given realization of the disorder,
can be solved numerically exact by using quite involved optimization algorithms.
Instead of using such a complex algorithmic approach that 
takes into account \emph{global} 
information we here address the question how well local algorithms perform.
 In particular we investigate
how well the transition point 
$\rho_c=0.0869(2)$
for the $2D$ square lattice with bimodal disorder,
resulting from \emph{static} simulations considering the NWP model, can be resolved 
using simple dynamic algorithms. We study local algorithms because
the movement of particles in physical systems, or the decisions of
agents in populations, typically depend on local information only. Hence, such an approach reflects real situations better.
Therefore we consider the dynamic biased and greedy RW dynamics discussed previously
in Sect.\ \ref{subsect:BGRW} and the ant colony optimization 
heuristic described in 
Sect.\ \ref{subsect:ACO}. The results are reported in 
Sects.\ \ref{subsect:res_BGRW} and \ref{subsect:res_ACO}, respectively.

\subsection{Biased and greedy random walks} \label{subsect:res_BGRW}

\begin{figure}[t!]
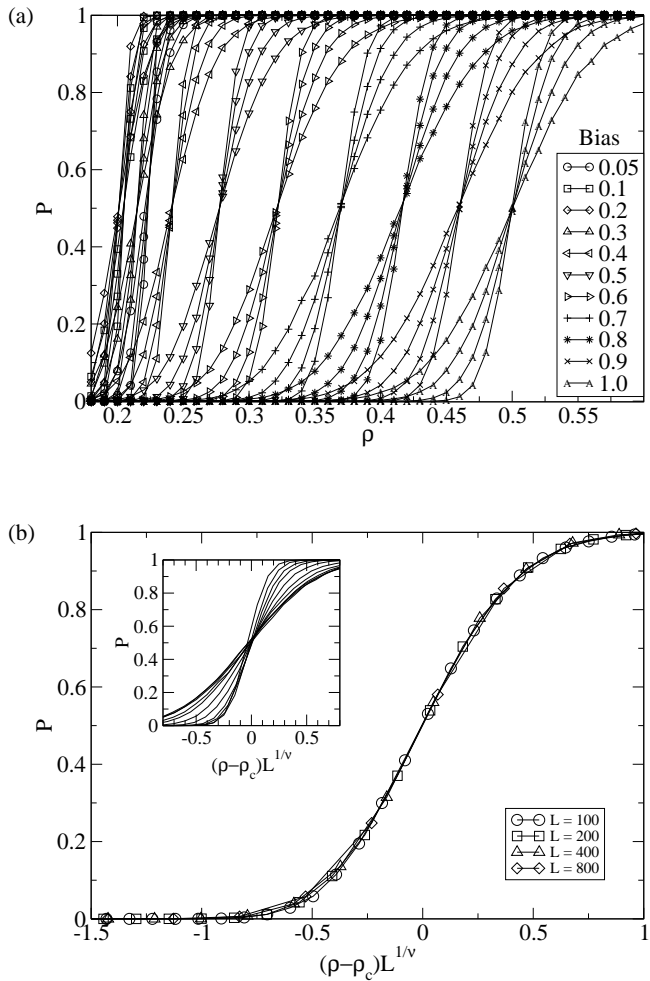

\begin{center}
\includegraphics[width=1.0\linewidth, bb= 10 30 730 600]{./P_A_both_b.eps}
\includegraphics[width=1.0\linewidth, bb= 10 30 730 600]{./P_A_both_b_rescaled_0_5.eps}
\end{center}
\caption{Results for type {\rm A} BGRWs where a walker interacts with the 
lattice graph according to rule (i) described in the text. The figure
shows the probability $P(\rho)$ of finding a lattice walk with a negative 
weight as function of the disorder parameter $\rho$.
(a) $P(\rho)$ for different bias strengths $B$ (listed in the key).
For a given value of $B$ the respective $4$-tuple of curves correspond
to the system sizes $L=100, 200, 400, 800$ (the larger the system size the steeper the curve).
The common crossing point indicates a critical value $\rho_c(B)$ at which,
in the thermodynamic limit, lattice walks with negative energy might
be found.
(b) the main plot shows a scaling plot for the particular 
bias strength $B=0.5$, therein the abscissa was 
rescaled according to $(\rho-\rho_{\rm c})L^{1/\nu}$ with $\rho_{\rm c}=0.2777(6)$
and $\nu=1.97(4)$.
The inset illustrates (for fixed $L=800$) the increasing steepness
of data curves for a decreasing bias parameter $B$.
I.e., the steepest curve is found for $B=0.05$, while at $B=1.0$ the
curve increases most gradual.
\label{fig:results_BGRW_typeA}}
\end{figure}  

At first, we consider type {\rm A} BGRWs where the walker interacts with
the environment via rule (i), i.e.\ while traversing the graph it
replaces positive and negative edge weights by edge weights equal to zero.
In Fig.\ \ref{fig:results_BGRW_typeA} we show the probability that,
after termination, a lattice walk with negative weight is found.
Simulations are carried out for systems of size $L=100, 200, 400, 800$
where data points represent an average over $2\times 10^4$ realizations
of the disorder.

To facilitate intuition, for a bias strength $B=1$ the greediness of the
walker has no effect. It starts at the outer node on the, say, left lattice boundary and performs
a straight line (i.e.\ $L+1$-step) walk to arrive at the outer node on 
the right lattice boundary (thus, the lattice walk is characterized by a
fractal dimension $d_f=1$). Since the disorder is uncorrelated, the 
walker basically sums up $L-1$ uncorrelated random edge weights drawn 
from the disorder distribution $P_1$, see Eq.\ (\ref{eq:disorderBimodal}).
At the value $\rho=0.5$ of the disorder parameter and for large values
of $L$ (on average) half of the edge weights will be negative. Hence, 
above $\rho_c(B=1)=0.5$, the value of $P(\rho)$ will tend to $1$ as
$L\to \infty$. At $\rho=0.5$ this is equivalent to a symmetric $1D$ RW,
where, starting at the origin, one allows the walker to perform $L-1$ 
steps, afterwards asking for the probability that the walker is located, say, left of the origin.
Also note that in this limit, the effective $1D$ problem statement is
trivially equivalent to the $1D$ NWP problem.
To further understand the scaling behavior of the data curves for the
case $B=1$ shown in Fig.\ \ref{fig:results_BGRW_typeA}, one might
follow a real-space renormalization approach 
\cite{wilson1971,wilsonNobelPrizeLecture1983,stanley1999}.
The basic idea is to replace a subsystem of $L$ edges (characterized
by disorder parameter $\rho$, which is, generally spoken,
 the probability to have
a negative weight) by one edge characterized via
an effective disorder parameter $\rho'=P(\rho)$.
In this regard, note that the only condition to yield a 
``successful'' negative weighted $L$-step path is that the number of
negative edge weights along the lattice walk needs to exceed the number 
of positive edge weights
by at least one. In this question, the probability to find a negative 
weighted path for systems of odd $L$ for a given value of $\rho$ is simply
a sum of binomial terms:
\begin{align}
P(\rho)=\sum_{k=L_0}^{L} \binom{L}{k}\rho^k(1-\rho)^{L-k}, \label{eq:probNegPath}
\end{align} 
where $L_0=(L+1)/2$. The fixed points $\rho^*$ for $P(\rho)$, which
charazterize phases/ phase transitions in the renormalization,
 are determined by $\rho^*=P(\rho^*)$.
Intuitively, a stability analysis of the three fixed points $\rho=0, 0.5, 1$ reveals
that the only unstable fixed point is the previously discussed value $\rho_c=0.5$ (at $B=1$).
Linearizing $P(\rho)$ about this critical point yields the scaling power $\nu$ of
the observable Eq.\ (\ref{eq:probNegPath}) via 
\begin{align}
\nu= \ln(L)/\ln(dP(\rho)/d\rho), \label{eq:Nu}
\end{align}
Albeit the expression for $\nu$ converges rather slowly, e.g.\ it assumes the values $\nu(L=10)=2.56$, $\nu(L=100)=2.22$ 
and $\nu(L=1000)=2.14$, it attains the limiting value $\nu=2$ as $L\to \infty$ (obtained through a
series expansion of the derivative of Eq.\ (\ref{eq:probNegPath}) in powers of $L$).
This slow convergence is presumably caused by the additive character of edge weights: the total sum 
of the edges must be negative, while sub-systems of arbitrary size are free to have a positive weight. 
Consequently, if the data curves for $P(\rho)$ obtained in the simulations 
exhibit the property of scaling it should thus not come as a surprise if the 
scaling behavior is governed by the scaling parameter $\nu=2$. In deed, 
we find that the data curves show a smooth data collapse if rescaled 
according to 
\begin{align}
P(\rho) = f[(\rho-\rho_c)L^{1/\nu}], \label{eq:scalingP}
\end{align}
where for $B=1$ the parameter values $\rho_c=0.5$ and $\nu=2$ yield a good
data collapse (not shown). This scaling behavior still holds 
for values of $B<1$, where $\rho_c<0.5$. E.g., in Fig.\ \ref{fig:results_BGRW_typeA} 
the scaled data for $B=0.5$ is illustrated, where $\rho_c=0.2777(6)$ and $\nu=1.97(4)$.

\begin{figure}[t!]
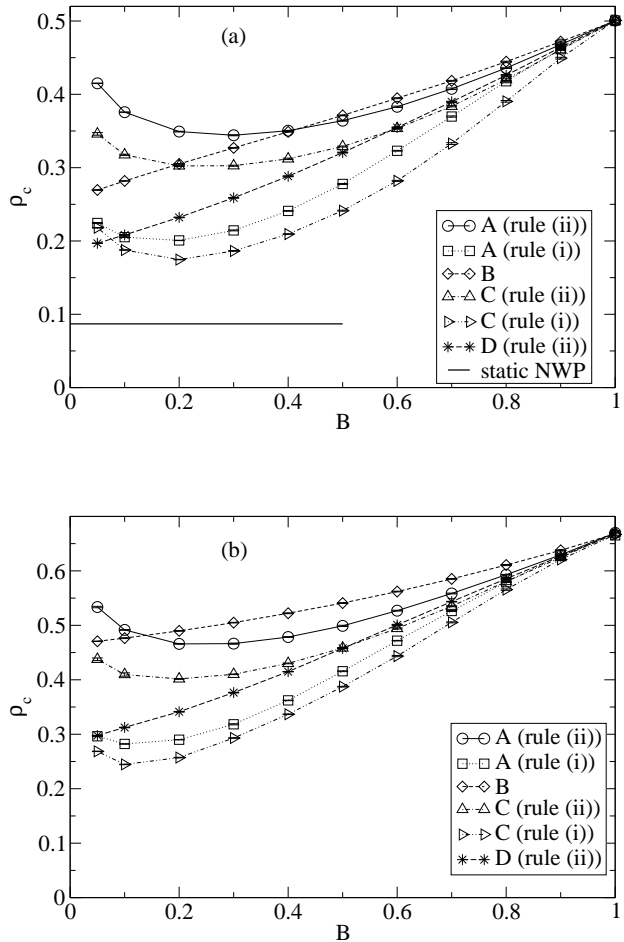

\begin{center}
\includegraphics[width=1.0\linewidth, bb= 10 30 740 600]{./Rho_b_MWPM_v2.eps}
\includegraphics[width=1.0\linewidth, bb= 10 30 740 600]{./Rho_f_v2.eps}
\end{center}
\caption{The critical threshold $\rho_c$ as a function of the bias parameter $B$ 
for the different types of BGRWs.
(a) Bimodal disorder drawn from Eq.\ (\ref{eq:disorderBimodal}). The continuous line represents 
the value of $\rho_c=0.0869(2)$ obtained using the static simulations in 
the context of the NWP problem. 
(b) Semi-continuous disorder drawn from Eq.\ (\ref{eq:disorderSemiContinuous}). 
\label{fig:results_BGRW_threshold}}
\end{figure}  

\begin{table}[b!]
\caption{\label{tab:tab1}
Values $B_{\rm c}$ of the bias strength for which the smallest $\rho_c$ is attained 
for all types of BGRWs considered.
The type of interaction with the graph is specified by rule (i) [rule (ii)] in the case where 
positive and negative [only negative] edge weights are modified during the walk.
The type of disorder distribution is indicated by ``bi'' (bimodal distribution; 
see Eq.\ \ref{eq:disorderBimodal}) 
and ``sc'' (semi-continuous; see Eq.\ \ref{eq:disorderSemiContinuous}).
In case of type {\rm B} and {\rm D} walks, the values of $\rho_c$ where estimated
from a fit of the data to a polynom of order three. We here list the figures up to
the third decimal, only (the estimated fit errors where
notoriously small; on the order of $10^{-6}$).}
\begin{ruledtabular}
\begin{tabular}[c]{l@{\quad}lllll}
BGRW type & $B_{\rm c}$ (bi) & $\rho_{\rm c}$ (bi) & $B_{\rm c}$ (sc) & $\rho_{\rm c}$ (sc) \\
\hline
A (rule (ii)) & 0.30(1) & 0.344(4) & 0.27(1) & 0.464(5) \\ 
A (rule (i)) & 0.18(1) & 0.199(4) & 0.12(1) & 0.280(3) \\
B     & 0.0         & 0.258        & 0.0         & 0.465   \\ 
C (rule (ii)) & 0.23(1) & 0.300(4) & 0.20(1) & 0.401(6) \\ 
C (rule (i)) & 0.18(1) & 0.174(6) & 0.12(1) & 0.241(4)\\ 
D     & 0.0         & 0.187       & 0.0         & 0.286 \\
\end{tabular}
\end{ruledtabular}
\end{table}

For decreasing bias strength $B<1$ the greediness of the BGRW has an
increasing influence on the dynamics of the walker. Hence, 
at a given value of $\rho$ the average path weight after a fixed 
number of steps is smaller at smaller bias strengths. Consequently
the probability to find a negative weighted path principally increases.
Albeit true for intermediate bias strengths $B=0.2 \ldots 1$, this
does not hold for comparatively small values $B<0.2$. As evident 
from Fig.\ \ref{fig:results_BGRW_threshold}, the critical point $\rho_c$
above which negative paths appear first as $L \to \infty$ tends to 
increase for a decreasing bias strength $<0.2$.
I.e., one can observe a minimum value of $\rho_{\rm c}=0.199(4)$ at a characteristic
value $B_{\rm c}\approx 0.18(1)$ (for type {\rm A} BGRWs considering an 
interaction with the environment according to rule (i)), see Tab.\ \ref{tab:tab1}.
A further extremal parameter set for type {\rm A} BGRWs is the case of 
vanishing bias strength $B=0$ at $\rho>0.5$. Since $\rho_{\rm c, bond}=0.5$ indicates
the onset of a system spanning (percolating) cluster of sites joined
by edges of weight $-1$, for $\rho>\rho_{\rm c, bond}$ there is a nonzero
probability that the walker crosses the lattice without accumulating a 
single positive edge weight. 

\begin{figure*}[t!]
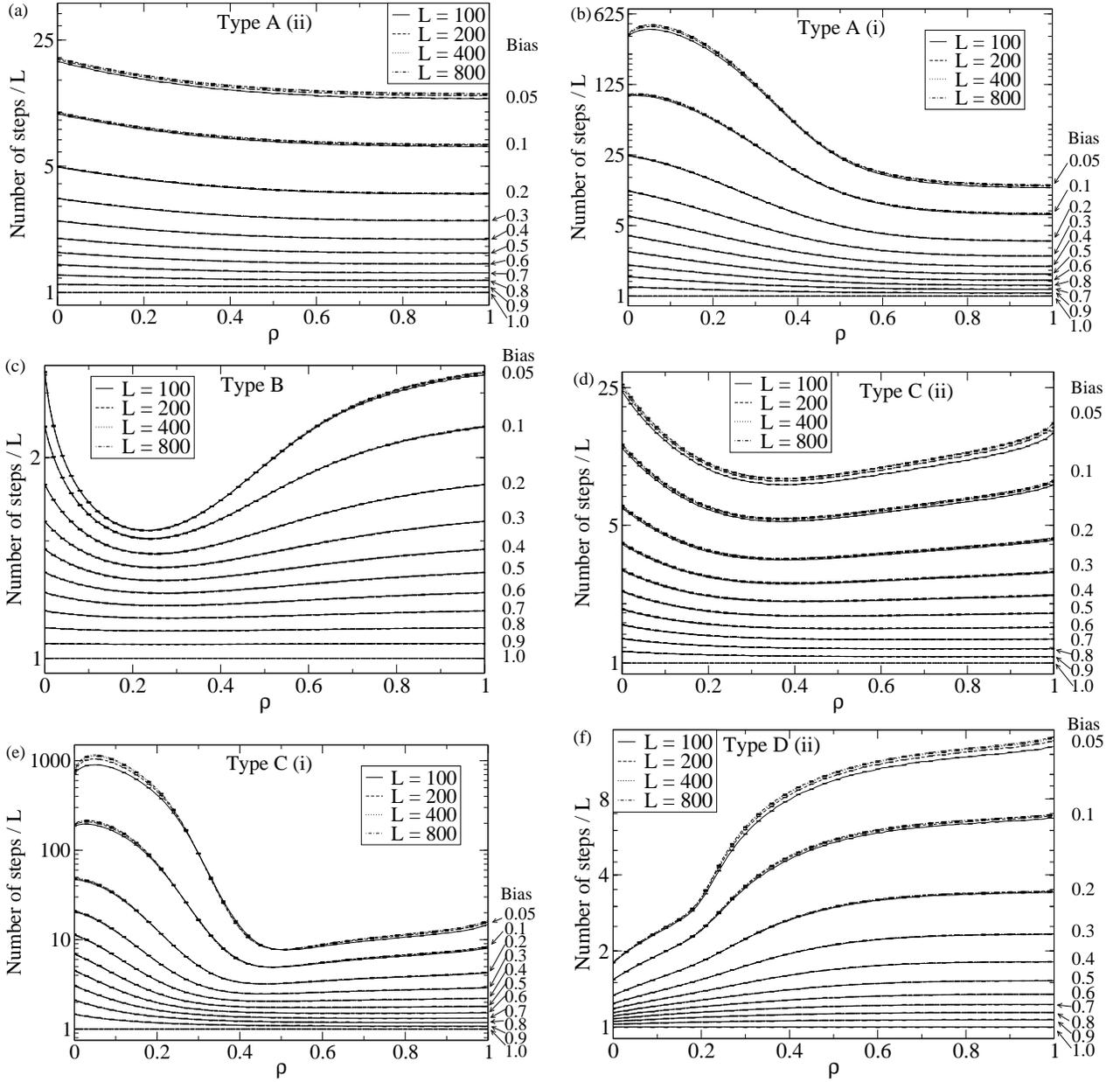

\begin{center}
\includegraphics[width=0.45\linewidth]{Steps_rescaled_A_neg_b}\hspace*{0.02\linewidth}
\includegraphics[width=0.45\linewidth]{Steps_rescaled_A_both_b}
\includegraphics[width=0.45\linewidth]{Steps_rescaled_B_b} \hspace*{0.02\linewidth}
\includegraphics[width=0.45\linewidth]{Steps_rescaled_C_neg_b}
\includegraphics[width=0.45\linewidth]{Steps_rescaled_C_both_b} \hspace*{0.02\linewidth}
\includegraphics[width=0.45\linewidth]{Steps_rescaled_D_neg_b}

\end{center}
\caption{Results for the average number of steps taken by 
the different BGRW types until the walk terminates. The figures (a) through (f)
show the average number of steps as function of the disorder parameter 
$\rho$ for the bimodal disorder distribution, 
for different bias strengths $B$ (see right border
of figures, respectively) and for different system 
sizes $L$. Note that, because the number of steps
is normalized by the system size $L$, the system-size dependence is very
small, the main differences are through the bias parameter $B$.
\label{fig:results_BGRW_Nsteps}}
\end{figure*}  

Regarding BGRW types {\rm A} through {\rm D}, Fig.\ \ref{fig:results_BGRW_threshold} shows
the critical points $\rho_c$ as function of bias strength $B$ for both, the bimodal disorder
distribution Eq.\ (\ref{eq:disorderBimodal})
 (see Fig.\ \ref{fig:results_BGRW_threshold}(a))
and the semi-continuous disorder Eq.\ (\ref{eq:disorderSemiContinuous})
 (see Fig.\ \ref{fig:results_BGRW_threshold}(b)).
As evident from the figure, BGRWs of type {\rm A} and {\rm C} show a 
minimum at critical disorder parameter value $\rho_{\rm c}$ at bias strengths in the range $B=0 \ldots 0.3$.
In contrast to this, BGRWs of type {\rm B} and {\rm D} appear to reach a minimum only 
when the bias strength is zero. 
This observation holds for both, bimodal and semi-continuous disorder, although for the different
disorder types the minima 
are located at different values $\rho_{\rm c}$ and $B_{\rm c}$.
For all the types of BGRWs considered, Tab.\ \ref{tab:tab1} lists the lowest value of $\rho_c$, 
attained at the corresponding bias strength $B_{\rm c}$. This can be thought of as a measure of 
efficiently for the different walk dynamics.
The lower the dynamical threshold $\rho_c$, the more
 successful a walker upon traversing
 a graph aimed at finding a path of total negative weight (under the respective dynamics).

Another quantity that allows to characterize the different dynamics is the average number of steps taken
by the different BGRW types until the walk terminates, see Fig.\ \ref{fig:results_BGRW_Nsteps}. This
quantity can further be used to determine the fractal scaling 
dimension of the walks: given that a 
walk of $N$ steps spans an linear distance $L$, its associated scaling dimension is defined via
$N\propto L^{d_{\rm f}}$. 
Note that for the resulting paths, due to the possibility
of loop erasure for some algorithms, the fractal dimension might be different.
On a general basis, as soon as there is a bias that causes an effective drift 
of a walker (as in the BGRWs considered here), one can expect that on large enough lattices one
trivially finds $d_{\rm f}=1$. Note that this is different when the environment itself is fractal, as 
for biased RWs on percolation clusters discussed in sect.\ \ref{sect:literatureReview}.
Here, for completeness and to illustrate the precise dependence of the dynamics on the disorder parameter
$\rho$ and different bias strengths $B$, the rescaled number of steps $N/L$ taken during the lattice walks
under the four BGRW dynamics are shown in Fig.\ \ref{fig:results_BGRW_Nsteps}
for the bimodal disorder distribution.

Considering type {\rm A} BGRWs, see Figs. \ref{fig:results_BGRW_Nsteps}(a) and (b), it can be seen that 
the number of steps generally decreases when the density of negative edges is increased. This can be easily 
explained by taking into account the greedy behavior of the walker. 
In case that the walker only consumes the negative edges (replacing them with a standard positive value of 1; 
see rule (ii) discussed above), this behavior appears since, after each step taken, the probability of 
returning to the previous node is smaller than that of proceeding forward towards other directions that might have 
negative edges. In the case where a type {\rm A} BGRW explores a graph with a low value of the disorder parameter 
and alters both, positive and negative edge weights (replacing them by edges with weight zero), an effective ``repulsive effect''
takes place because of the comparatively large probability that it might encounter excess edges with positive weights.
Hence, the walker is likely to return to its previous location by traversing the edge with weight zero created 
while performing the last step. This repulsive effect is emphasized especially for a bias strength $B<0.5$. 
As can be seen in Fig.\ \ref{fig:results_BGRW_Nsteps}(b), this repulsive effect leads to an increase in the number 
of steps taken by more than one order of magnitude in the low bias regime, compared to the case when only the 
negative bonds are altered. 
Also note that for different system sizes $L$ the data curves for the scaled number of steps $N/L$ fall on top of
each other. Hence, the expected scaling dimension $d_{\rm f}=1$ can be verified from the figures.

Regarding BGRWs of type {\rm B}, illustrated in Fig.\ \ref{fig:results_BGRW_Nsteps}(c), the number of steps 
taken in the extreme cases $\rho=0$ (positive edge weights only) and $\rho=1$ (negative edge weights only),
where there is no disorder at all, is the same, as one would expect.
Regarding the bimodal disorder distribution, a minimum in the walk length is found when the disorder parameter assumes the value $\rho\approx 0.25$.
This is caused by the fact that each node is, on average, endnode of one edge with negative edge weight that acts 
like a local trap for the walker. Only under the influence of the bias the walker can escape from such a trap.
Until then, the walker has possibly traversed the edge with negative weight multiple times.
After the loop erasure process (discussed for type {\rm B} BGRWs in subsect.\ \ref{subsect:BGRW}), this results
in an almost straight-line walk.
Increasing $\rho$ above $\rho=0.25$ also increases the likelihood of a node of having, on average, 
more than one incident edge with negative edge weight, thus preventing the walker from getting stuck
(however, note that for the $2D$ setup considered here, only above $\rho=0.5$ there is a system spanning
cluster of negative edge weights).
Although not shown here, in the case of a semi-continuous disorder distribution, the number of steps shows a similar behavior up 
to $\rho=0.25$ but instead of increasing afterwards it has a steady decrease until it reaches $\rho=1$
density which is caused by the lack of degeneracy in the bond weights. 

A close similarity between BGRWs of type {\rm A} and {\rm C} can be seen upon comparison of the respective
subfigures in Fig.\ \ref{fig:results_BGRW_Nsteps}. The most evident difference between the two dynamics
is that type {\rm C} BGRWs do not reach the minimum number of steps when the graph is fully equipped with negative edge weights.
This effect appears due to the additional optimization overhead that takes advantage of the larger number of negative
 edge weights at large values of $\rho$ by 
increasing the length of the random walk. 

For type {\rm D} BGRWs, shown in Fig.\ \ref{fig:results_BGRW_Nsteps}(f), a behavior strikingly different from the other three 
BGRW dynamics can be observed. The respective lattice walks exhibit a minimum in the number of steps 
when no negative edges are present in the graph and attain a maximum when the graph is 
fully equipped with negative edge weights. This is a tell tale sign of an increase in the efficiency of traversing edges
with a negative weight, while avoiding positive edge weights. Shorter lattice walks are performed at small values of $\rho$ and
longer paths, possibly with many detours, are found at larger values of $\rho$. The latter characteristic for the optimal paths
in a graph predominantly equipped with negative edge weights.

	It is interesting to note that the more efficient a random walk algorithm is at finding negative weighted paths, the 
higher the number of steps it will take at a low density of negative bonds if it alters both, negative and positive bonds,
corresponding to rule (i) discussed above. 
As we observe, this effect is especially pronounced in the case of BGRWs of type {\rm D}. The respective algorithm could not even be used 
efficiently for $\rho<0.5$ (no results for this case are shown).

As pointed out above, the expected scaling dimension for the BGRWs is $d_{\rm f}=1$. 
The number of steps taken during the lattice walks, normalized by the system size
verifies this in the form $\langle N \rangle/L\approx {\rm const}$ for given values of $\rho$ and $B$.
Discrepancies in the curve overlap, corresponding to deviations from the expected scaling
behavior, are accentuated at small values of $B$, because the size of the walk gets closer 
to a $B$ dependent crossover length $L_c$ (similar to the crossover time $\tau_c$ discussed in sect.\ \ref{sect:literatureReview}). 
For walks much shorter than this crossover length the BGRWs exhibit a scaling dimension 
larger than $1$. Only for walks longer than the crossover length, they assume the asymptotic
scaling dimension $d_{\rm f}=1$. The simple explanation for this effect is that on short
length scales the diffusive (albeit greedy) behavior dominates the configuration of the 
lattice walk, while on longer lengthscales the drift induced by the bias is dominant.
Note that the scaling dimension of the lattice walks obtained here is very different from the
fractal scaling dimension of the paths in the $2D$ NWP problem.
For the $2D$ setup of the NWP problem, a fractal scaling dimension of $d_{\rm f}=1.268(1)$ was observed at $\rho=0.1032(5)$ \cite{melchert2008},
increasing towards $d_{\rm f}=1.756(8)$ at $\rho=1$ (measured for the loops-only setup) \cite{melchert2011a}.


\subsection{Ant colony optimization} \label{subsect:res_ACO}

\begin{figure}[t!]
\begin{center}
\includegraphics[width=1.0\linewidth]{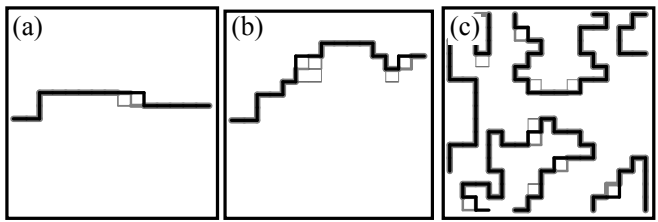}
\end{center}
\caption{Exemplary $(s,t)-$paths found by the ACO algorithm for different fractions $\rho$
of negative edge weights on a $2D$ square lattice of side length $L=16$. 
(a) $\rho=0.10$, (b) $\rho=0.15$, and (c) $\rho=0.90$. 
The black path indicates
the best path found after the algorithm terminated and the line-width of 
the other edges (colored grey) indicate how frequently the respective edge 
was visited by an agent.
\label{fig:nwpAco_examples}}
\end{figure}  

In order to find values of $\alpha$ and $\beta$ that reflect the underlying edge-weights,  
leading to an exploration in the vicinity of the ``best'' $(s,t)-$path found so far (referred to 
as \emph{intensification}) but still 
allowing for an efficient exploration of many different paths (referred to as \emph{diversification}), 
we performed several ``calibration''-runs of the ACO algorithm.
During our simulation we used $M=L$, $\alpha=1/M$, and $\beta=0.98$. Below we report 
the results of a finite-size scaling analysis for the probability that the ACO algorithm
finds a negatively weighted optimal path for different values of the fraction $\rho$ of negative
edge-weights on square lattice graphs with $L=8,\dots,20$. At each value of $\rho$, we considered 
$n=10^3$ different realizations of the edge weights drawn from a bimodal disorder distribution (Eq.\ \ref{eq:disorderBimodal}) in order to compute averages.
The considered systems have periodic boundaries in the, say, vertical direction and open boundaries
in the horizontal direction. Further, two extra ``outer'' nodes, i.e.\ the source node $s$ and the 
target node $t$, are introduced. The source (target) node is connected to all nodes on the left (right) system boundary as shown in Fig.\ \ref{fig:latticeWalk}.
Consequently, a $(s,t)-$path spans the system along the horizontal direction. Depending on the 
fraction $\rho$ of negative edge weights it might also wrap the system along the vertical direction. 
Fig.\ \ref{fig:nwpAco_examples} illustrates exemplary $(s,t)-$paths found by the ACO algorithm 
for three different values of the disorder parameter on a $2D$ square lattice of side length $L=16$. 
In the figure, the black path indicates the best path found after the algorithm terminated 
and the line-width of the other edges (colored grey) indicate how frequently the respective edge 
was visited by an agent.
As evident from Fig.\ \ref{fig:nwpAco_examples}(a), corresponding to $\rho=0.10$, if there are only few 
negative edge weights, the ``best'' $(s,t)-$path found by the ACO algorithm is rather straight-lined.
As the fraction of negative edge weights increases, see Figs.\ \ref{fig:nwpAco_examples}(b,c), corresponding
to $\rho=0.15$ and $\rho=0.90$, respectively, the paths roughen up and eventually also wind around the 
lattice  in the vertical direction.

\begin{figure}[t!]
\begin{center}
\includegraphics[width=1.0\linewidth]{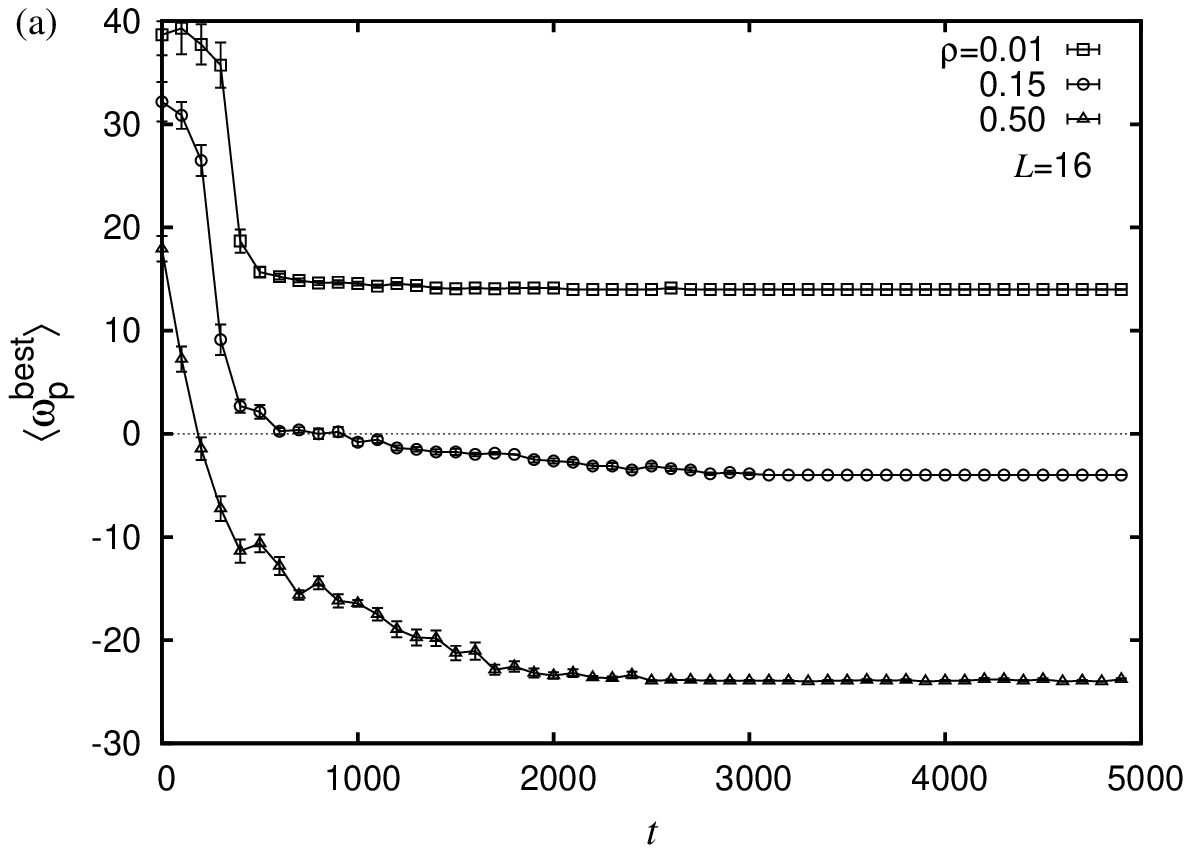}
\includegraphics[width=1.0\linewidth]{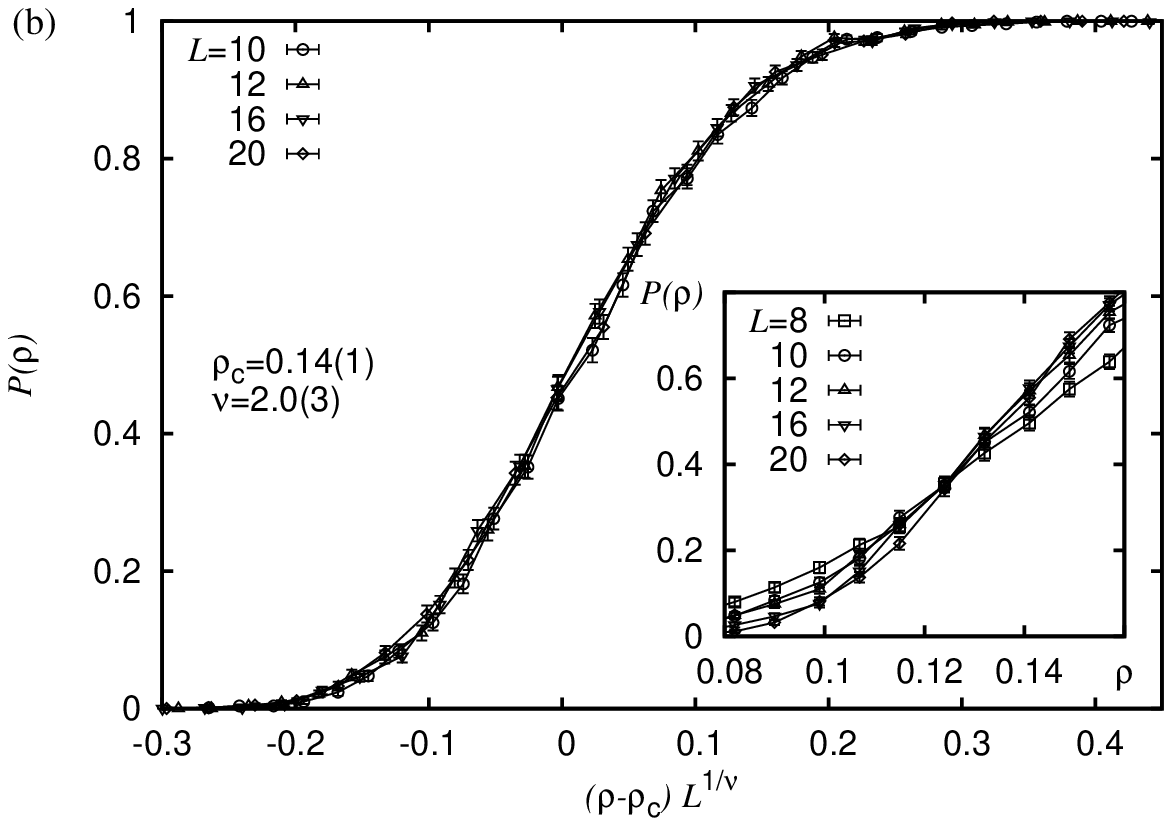}
\end{center}
\caption{Results for the ACO heuristic on $2D$ square lattice graphs
for the case of a bimodal disorder distribution.
(a) Best path-weight found so far as function of the number of sweeps 
carried out by the ACO algorithm, averaged over 
different realizations of the edge weight disorder at $L=16$ for three
different values of $\rho$.
(b) Probability $P(\rho)$ that the ACO algorithm returns a $(s,t)-$path with 
negative weight as a function of the fraction $\rho$ of negative edge weights.
The main plot shows the scaled data (see text for more details)
and the inset illustrates the unscaled data in the vicinity of 
the critical point $\rho_{\rm c}=0.14(1)$.
\label{fig:nwpAco_results}}
\end{figure}  

The characteristics of the weight of the best path found by the ACO algorithm as a function of the 
number of sweeps carried out by the algorithm is shown in Fig.\ \ref{fig:nwpAco_results}(a).
As can be seen in the curve corresponding to $\rho=0.01$ and after a short exploration time ($t=0-500$ sweeps), 
the ACO algorithm maintains an average best pathweight $\langle\omega_{\rm p}^{\rm best}\rangle\approx L$.
Since the underlying edge-weight distribution is bimodal, i.e.\ it allows for $\omega_{ij}=\pm 1$ only, 
this reflects that at very small values of $\rho$ almost all edges contained in $(s,t)-$paths will have
weight $1$ and the paths tend to have length $\approx L$.
The situation changes upon increasing the value of $\rho$. As can be seen from Fig.\ \ref{fig:nwpAco_results}(a),
already at $\rho=0.15$ and after a certain initial exploration time ($t=0-1000$), the ACO algorithm is 
able to identify $(s,t)-$paths with a negative weight so that $\langle\omega_{\rm p}^{\rm best}\rangle<0$. 

Finally, Fig.\ \ref{fig:nwpAco_results}(b) illustrates the scaling behavior of the probability $P(\rho)$, 
that the ACO algorithm returns a negative weighted $(s,t)-$path as a function of the fraction $\rho$ of 
negative edge weights.
As can be seen from the inset of Fig.\ \ref{fig:nwpAco_results}(b), the data curves for the different
system sizes cross at $\rho_c\approx 0.13$. Below (above) $\rho_c$ and for increasing system size, the probability 
that the ACO algorithm returns a path weight $\omega_{p}<0$ tends to zero (one).
As can be seen from the main plot of Fig.\ \ref{fig:nwpAco_results}(b), $P(\rho)$ can be rescaled using
Eq.\ (\ref{eq:scalingP}), similar to the paths found using the BGRW dynamics discussed previously.
Considering the system sizes $L\geq 10$, a best data collapse (attained in the range $(\rho-\rho_{\rm c})L^{1/\nu}\in[-0.3,0.3]$ on 
the rescaled $\rho$-axis) results in the estimates $\rho_{\rm c}=0.14(1)$ and $\nu=2.0(3)$ with a quality $S=1.84$ of 
the data collapse \cite{autoScale2009} (the numerical value of $S$ measures the mean--square distance of the data points to the master
curve, described by the scaling function, in units of the standard error
\cite{houdayer2004}).
Performing the analysis for different intervals on the rescaled $\rho$-axis led to various estimates in the range $\rho_{\rm c}=0.13 \ldots 0.14$. 
Further estimates of $\rho_{\rm c}$ and $\nu$ can be obtained 
from the scaling behavior of the variance ${\rm var}(P)$. It assumes a peak at an effective, $L$-dependent 
critical point $\rho_{L}^{\rm eff}$ at which $P(\rho_L^{\rm eff})=0.5$ (not shown). We find that $\rho_L^{\rm eff}$ 
seems to saturate at a value consistent with $\rho_{\rm c}$. In particular, we find $\rho_{20}^{\rm eff}=0.137(1)$ 
(for the smaller system size $L=10$ we find $\rho_{10}^{\rm eff}=0.137(2)$), obtained by fitting a Gauss-shaped 
curve to the data of ${\rm var}(P)$. The value of $\rho_L^{\rm eff}$ specifies the location of the peak of 
the fitting-curve and the error-bar is the respective fitting error. 
Further, the width $\sigma_L$ of the Gauss-shaped fitting function is consistent with 
$\sigma_L=a L^{-1/\nu}$, where we found $a=0.098(6)$ and $\nu=2.1(1)$ considering $L=8\ldots20$.
Note that the value of the exponent $\nu$ found here is also in agreement with the one found for the BGRWs in 
the preceeding subsection.


\section{Conclusions}\label{sect:conclusions}

In the presented article we have investigated the principle characteristics
of biased greedy random walks (BGRWs) on $2D$ lattices with quenched disorder on 
the lattice edges. Four different types of BGRWs and an algorithm based on the 
ant colony optimization (ACO) heuristic were considered.
Regarding the BGRWs, the precise configurations of the lattice walks constructed 
during the numerical simulations were influenced by two parameters: 
a disorder parameter $\rho$ that controls the amount of negative edge weights 
on the lattice and a bias strength $B$ that governs the drift of the walkers 
along a certain lattice direction.
Focus of the presented study was the probability that, after termination, a lattice walk
exhibits a negative weight as function of $\rho$ and $B$.
All four types of dynamics exhibit a phase
transition with a characteristic disorder value $\rho_{\rm c}$ above which 
lattice walks with a negative weight appear first. Further, the data curves for the 
probability that a walk with negative energy is found as function of $\rho$ could be rescaled
well according to a simple scaling form. The respective scaling parameters in the extremal case
of maximal bias $B=1$, where the dynamics is effectively one dimensional, could be estimated from 
a simple renormalization argument. A lower bound on the value of $\rho_{\rm c}$
can be obtained from the NWP problem in $2D$. Regarding those ``static'' simulations considering 
bimodal disorder on $2D$ square lattices, negative-weighted paths appear first above the 
threshold $\rho_{\rm c}=0.0869(2)$ with critical exponent
$\nu = 1.5(1)$.
Considering the BGRW dynamics only, the 
type {\rm C} BGRW (implementing rule (i) for the walker--edge weight 
interaction discussed in 
sect.\ \ref{subsect:BGRW}) yields the closest estimate, reading $\rho_{\rm c}=0.174(6)$ at bias strength $B_{\rm c}=0.18(1)$,
followed closely by type {\rm D} lattice walks with $\rho_{\rm c}=0.18(1)$ at vanishing bias, see Tab.\ \ref{tab:tab1}.
Correspondingly, algorithm A (i) needs the largest number of steps to 
yield the best results. Interestingly, algorithm D is considerably faster,
also compared to almost all variants. 

The ACO based algorithm outperforms these estimates, yielding $\rho_{\rm c}=0.14(1)$. 
Note that, in comparison to the type {\rm A} and {\rm B} BGRWs, which perform slightly worse,
the dynamic rules governing the type {\rm C} and {\rm D} lattice walks and the ACO heuristic 
are rather involved.
Also note that, albeit all the dynamic rules implement some degree of ``local'' optimization, 
they were not able to reach the threshold value obtained from the static simulations in terms of the NWP problem. Also the value $\nu=2$ of
the critical exponent, for all types of BGRW and ACO dynamics, 
and the fractal behavior of the walks, hence of
the paths, is different from the static negative weight paths. Thus
static and dynamic behavior of the problem is different, as for many
glassy systems.

Whether one can reach this static threshold and find the
same critical exponents by means of simple dynamics 
that implement local optimization 
remains an intriguing open problem. 
This could be possible in principle, because the
corresponding optimization problem can be solved in polynomial time.
At least we could not succeed in 
doing so via the 
biased greedy random walks considered in the presented study. 

Another interesting problem is whether a collective algorithm, which
explores, e.g., the dynamics of an ensembles of paths connected to a heat
bath at temperature $T$, 
with the path weight being the energy, exhibits phase transitions
as function of the disorder parameter $\rho$ as well. 
These transitions could be in the limit $T\to 0$ 
similar to the phase transitions 
found previously for the exact algorithms. For lower temperatures,
maybe below a critical dynamical threshold, these transitions could
be more similar to
the transitions investigated in this work.
Related studies are currently being performed.
	
\begin{acknowledgments}
TLM acknowledges support from ESF through the project POSDRU 107/1.5/S/80765.
OM acknowledges financial support from the DFG (\emph{Deutsche Forschungsgemeinschaft}) 
under grant HA3169/3-1. 
The simulations were performed at the HPC Cluster HERO, located at 
the University of Oldenburg (Germany) and funded by the DFG through
its Major Instrumentation Programme (INST 184/108-1 FUGG) and the
Ministry of Science and Culture (MWK) of the Lower Saxony State.
\end{acknowledgments}


\bibliography{nwp_dynamicApproach.bib}

\begin{thebibliography}{49}
\expandafter\ifx\csname natexlab\endcsname\relax\def\natexlab#1{#1}\fi
\expandafter\ifx\csname bibnamefont\endcsname\relax
  \def\bibnamefont#1{#1}\fi
\expandafter\ifx\csname bibfnamefont\endcsname\relax
  \def\bibfnamefont#1{#1}\fi
\expandafter\ifx\csname citenamefont\endcsname\relax
  \def\citenamefont#1{#1}\fi
\expandafter\ifx\csname url\endcsname\relax
  \def\url#1{\texttt{#1}}\fi
\expandafter\ifx\csname urlprefix\endcsname\relax\def\urlprefix{URL }\fi
\providecommand{\bibinfo}[2]{#2}
\providecommand{\eprint}[2][]{\url{#2}}

\bibitem[{\citenamefont{de~Gennes}(1971)}]{DeGennes1971}
\bibinfo{author}{\bibfnamefont{P.~G.} \bibnamefont{de~Gennes}},
  \bibinfo{journal}{La Recherche} \textbf{\bibinfo{volume}{7}},
  \bibinfo{pages}{919} (\bibinfo{year}{1971}).

\bibitem[{\citenamefont{Straley}(1980)}]{straley1980}
\bibinfo{author}{\bibfnamefont{J.~P.} \bibnamefont{Straley}},
  \bibinfo{journal}{J. Phys. C} \textbf{\bibinfo{volume}{13}},
  \bibinfo{pages}{2991} (\bibinfo{year}{1980}).

\bibitem[{\citenamefont{Stauffer}(1999)}]{Stauffer1999}
\bibinfo{author}{\bibfnamefont{D.}~\bibnamefont{Stauffer}},
  \bibinfo{journal}{Physica A} \textbf{\bibinfo{volume}{266}},
  \bibinfo{pages}{35 } (\bibinfo{year}{1999}).

\bibitem[{\citenamefont{Melchert and Hartmann}(2008)}]{melchert2008}
\bibinfo{author}{\bibfnamefont{O.}~\bibnamefont{Melchert}} \bibnamefont{and}
  \bibinfo{author}{\bibfnamefont{A.~K.} \bibnamefont{Hartmann}},
  \bibinfo{journal}{New. J. Phys.} \textbf{\bibinfo{volume}{10}},
  \bibinfo{pages}{043039} (\bibinfo{year}{2008}).

\bibitem[{\citenamefont{Apolo et~al.}(2009)\citenamefont{Apolo, Melchert, and
  Hartmann}}]{apolo2009}
\bibinfo{author}{\bibfnamefont{L.}~\bibnamefont{Apolo}},
  \bibinfo{author}{\bibfnamefont{O.}~\bibnamefont{Melchert}}, \bibnamefont{and}
  \bibinfo{author}{\bibfnamefont{A.~K.} \bibnamefont{Hartmann}},
  \bibinfo{journal}{Phys. Rev. E} \textbf{\bibinfo{volume}{79}},
  \bibinfo{pages}{031103} (\bibinfo{year}{2009}).

\bibitem[{\citenamefont{Melchert et~al.}(2010)\citenamefont{Melchert, Apolo,
  and Hartmann}}]{melchert2010a}
\bibinfo{author}{\bibfnamefont{O.}~\bibnamefont{Melchert}},
  \bibinfo{author}{\bibfnamefont{L.}~\bibnamefont{Apolo}}, \bibnamefont{and}
  \bibinfo{author}{\bibfnamefont{A.~K.} \bibnamefont{Hartmann}},
  \bibinfo{journal}{Phys. Rev. E} \textbf{\bibinfo{volume}{81}},
  \bibinfo{pages}{051108} (\bibinfo{year}{2010}).

\bibitem[{\citenamefont{Melchert et~al.}(2011)\citenamefont{Melchert, Hartmann,
  and M\'ezard}}]{melchert2011b}
\bibinfo{author}{\bibfnamefont{O.}~\bibnamefont{Melchert}},
  \bibinfo{author}{\bibfnamefont{A.~K.} \bibnamefont{Hartmann}},
  \bibnamefont{and} \bibinfo{author}{\bibfnamefont{M.}~\bibnamefont{M\'ezard}},
  \bibinfo{journal}{Phys. Rev. E} \textbf{\bibinfo{volume}{84}},
  \bibinfo{pages}{041106} (\bibinfo{year}{2011}).

\bibitem[{\citenamefont{Norrenbrock et~al.}(2012)\citenamefont{Norrenbrock,
  Melchert, and Hartmann}}]{Norrenbrock2012}
\bibinfo{author}{\bibfnamefont{C.}~\bibnamefont{Norrenbrock}},
  \bibinfo{author}{\bibfnamefont{O.}~\bibnamefont{Melchert}}, \bibnamefont{and}
  \bibinfo{author}{\bibfnamefont{A.~K.} \bibnamefont{Hartmann}}
  (\bibinfo{year}{2012}), \bibinfo{note}{{preprint: arXiv:1205.1412}}.

\bibitem[{\citenamefont{Claussen et~al.}(2012)\citenamefont{Claussen, Apolo,
  Melchert, and Hartmann}}]{claussen2012}
\bibinfo{author}{\bibfnamefont{G.}~\bibnamefont{Claussen}},
  \bibinfo{author}{\bibfnamefont{L.}~\bibnamefont{Apolo}},
  \bibinfo{author}{\bibfnamefont{O.}~\bibnamefont{Melchert}}, \bibnamefont{and}
  \bibinfo{author}{\bibfnamefont{A.~K.} \bibnamefont{Hartmann}},
  \bibinfo{journal}{Phys. Rev. E} \textbf{\bibinfo{volume}{86}},
  \bibinfo{pages}{056708} (\bibinfo{year}{2012}).

\bibitem[{\citenamefont{Young}(1998)}]{young1998}
\bibinfo{editor}{\bibfnamefont{A.~P.} \bibnamefont{Young}}, ed.,
  \emph{\bibinfo{title}{Spin glasses and random fields}}
  (\bibinfo{publisher}{World Scientific}, \bibinfo{address}{Singapore},
  \bibinfo{year}{1998}).

\bibitem[{\citenamefont{Hartmann and Weigt}(2005)}]{phase-transitions2005}
\bibinfo{author}{\bibfnamefont{A.~K.} \bibnamefont{Hartmann}} \bibnamefont{and}
  \bibinfo{author}{\bibfnamefont{M.}~\bibnamefont{Weigt}},
  \emph{\bibinfo{title}{Phase Transitions in Combinatorial Optimization
  Problems}} (\bibinfo{publisher}{Wiley-VCH}, \bibinfo{address}{Weinheim},
  \bibinfo{year}{2005}).

\bibitem[{\citenamefont{Binder and Kob}(2011)}]{binder2011}
\bibinfo{author}{\bibfnamefont{K.}~\bibnamefont{Binder}} \bibnamefont{and}
  \bibinfo{author}{\bibfnamefont{W.}~\bibnamefont{Kob}},
  \emph{\bibinfo{title}{Glassy Materials and Diosordred Solids}}
  (\bibinfo{publisher}{World Scientific}, \bibinfo{address}{Singapore},
  \bibinfo{year}{2011}).

\bibitem[{\citenamefont{Ben-Avraham and Havlin}(1982)}]{benavraham1982}
\bibinfo{author}{\bibfnamefont{D.}~\bibnamefont{Ben-Avraham}} \bibnamefont{and}
  \bibinfo{author}{\bibfnamefont{S.}~\bibnamefont{Havlin}},
  \bibinfo{journal}{J. Phys. A} \textbf{\bibinfo{volume}{15}},
  \bibinfo{pages}{L691} (\bibinfo{year}{1982}).

\bibitem[{\citenamefont{Gefen et~al.}(1983)\citenamefont{Gefen, Aharony, and
  Alexander}}]{gefen1983}
\bibinfo{author}{\bibfnamefont{Y.}~\bibnamefont{Gefen}},
  \bibinfo{author}{\bibfnamefont{A.}~\bibnamefont{Aharony}}, \bibnamefont{and}
  \bibinfo{author}{\bibfnamefont{S.}~\bibnamefont{Alexander}},
  \bibinfo{journal}{Phys. Rev. Lett.} \textbf{\bibinfo{volume}{50}},
  \bibinfo{pages}{77} (\bibinfo{year}{1983}).

\bibitem[{\citenamefont{Pandey}(1984{\natexlab{a}})}]{padey1984}
\bibinfo{author}{\bibfnamefont{R.~B.} \bibnamefont{Pandey}},
  \bibinfo{journal}{Phys. Rev. B} \textbf{\bibinfo{volume}{30}},
  \bibinfo{pages}{489} (\bibinfo{year}{1984}{\natexlab{a}}).

\bibitem[{\citenamefont{Argyrakis and Kopelman}(1984)}]{argyrakis1984No2}
\bibinfo{author}{\bibfnamefont{P.}~\bibnamefont{Argyrakis}} \bibnamefont{and}
  \bibinfo{author}{\bibfnamefont{R.}~\bibnamefont{Kopelman}},
  \bibinfo{journal}{J. Chem. Phys.} \textbf{\bibinfo{volume}{81}},
  \bibinfo{pages}{1015} (\bibinfo{year}{1984}), \bibinfo{note}{{A summary of
  this article is available at papercore.org, see
  {http://www.papercore.org/Argyrakis1984No2}}}.

\bibitem[{\citenamefont{White and Barma}(1984)}]{white1984}
\bibinfo{author}{\bibfnamefont{S.~R.} \bibnamefont{White}} \bibnamefont{and}
  \bibinfo{author}{\bibfnamefont{M.}~\bibnamefont{Barma}},
  \bibinfo{journal}{Journal of Physics A: Mathematical and General}
  \textbf{\bibinfo{volume}{17}}, \bibinfo{pages}{2995} (\bibinfo{year}{1984}).

\bibitem[{\citenamefont{Havlin and Ben-Avraham}(1987)}]{havlin1987}
\bibinfo{author}{\bibfnamefont{S.}~\bibnamefont{Havlin}} \bibnamefont{and}
  \bibinfo{author}{\bibfnamefont{D.}~\bibnamefont{Ben-Avraham}},
  \bibinfo{journal}{Advances in Physics} \textbf{\bibinfo{volume}{36}},
  \bibinfo{pages}{695} (\bibinfo{year}{1987}).

\bibitem[{\citenamefont{Bouchaud and Georges}(1990)}]{bouchaud1990}
\bibinfo{author}{\bibfnamefont{J.~P.} \bibnamefont{Bouchaud}} \bibnamefont{and}
  \bibinfo{author}{\bibfnamefont{A.}~\bibnamefont{Georges}},
  \bibinfo{journal}{Phys. Rep.} \textbf{\bibinfo{volume}{195}},
  \bibinfo{pages}{127 } (\bibinfo{year}{1990}).

\bibitem[{\citenamefont{Argyrakis et~al.}(1984)\citenamefont{Argyrakis,
  Anacker, and Kopelman}}]{argyrakis1984}
\bibinfo{author}{\bibfnamefont{P.}~\bibnamefont{Argyrakis}},
  \bibinfo{author}{\bibfnamefont{L.}~\bibnamefont{Anacker}}, \bibnamefont{and}
  \bibinfo{author}{\bibfnamefont{R.}~\bibnamefont{Kopelman}},
  \bibinfo{journal}{J. Stat. Phys.} \textbf{\bibinfo{volume}{36}},
  \bibinfo{pages}{579} (\bibinfo{year}{1984}), \bibinfo{note}{{A summary of
  this article is available at papercore.org, see
  {http://www.papercore.org/Argyrakis1984}}}.

\bibitem[{\citenamefont{Avramov et~al.}(1993)\citenamefont{Avramov, Milchev,
  and Argyrakis}}]{avramov1993}
\bibinfo{author}{\bibfnamefont{I.}~\bibnamefont{Avramov}},
  \bibinfo{author}{\bibfnamefont{A.}~\bibnamefont{Milchev}}, \bibnamefont{and}
  \bibinfo{author}{\bibfnamefont{P.}~\bibnamefont{Argyrakis}},
  \bibinfo{journal}{Phys. Rev. E} \textbf{\bibinfo{volume}{47}},
  \bibinfo{pages}{2303} (\bibinfo{year}{1993}), \bibinfo{note}{{A summary of
  this article is available at papercore.org, see
  {http://www.papercore.org/Avramov1993}}}.

\bibitem[{\citenamefont{Bunde et~al.}(1987)\citenamefont{Bunde, Harder, Havlin,
  and Roman}}]{bunde1987}
\bibinfo{author}{\bibfnamefont{A.}~\bibnamefont{Bunde}},
  \bibinfo{author}{\bibfnamefont{H.}~\bibnamefont{Harder}},
  \bibinfo{author}{\bibfnamefont{S.}~\bibnamefont{Havlin}}, \bibnamefont{and}
  \bibinfo{author}{\bibfnamefont{H.~E.} \bibnamefont{Roman}},
  \bibinfo{journal}{J. Phys. A} \textbf{\bibinfo{volume}{20}},
  \bibinfo{pages}{L865} (\bibinfo{year}{1987}).

\bibitem[{\citenamefont{Arapaki et~al.}(1997)\citenamefont{Arapaki, Argyrakis,
  Avramov, and Milchev}}]{arapaki1997}
\bibinfo{author}{\bibfnamefont{E.}~\bibnamefont{Arapaki}},
  \bibinfo{author}{\bibfnamefont{P.}~\bibnamefont{Argyrakis}},
  \bibinfo{author}{\bibfnamefont{I.}~\bibnamefont{Avramov}}, \bibnamefont{and}
  \bibinfo{author}{\bibfnamefont{A.}~\bibnamefont{Milchev}},
  \bibinfo{journal}{Phys. Rev. E} \textbf{\bibinfo{volume}{56}},
  \bibinfo{pages}{R29} (\bibinfo{year}{1997}), \bibinfo{note}{{A summary of
  this article is available at papercore.org, see
  {http://www.papercore.org/Arapaki1997}}}.

\bibitem[{\citenamefont{Avramov et~al.}(1998)\citenamefont{Avramov, Milchev,
  Arapaki, and Argyrakis}}]{avramov1998}
\bibinfo{author}{\bibfnamefont{I.}~\bibnamefont{Avramov}},
  \bibinfo{author}{\bibfnamefont{A.}~\bibnamefont{Milchev}},
  \bibinfo{author}{\bibfnamefont{E.}~\bibnamefont{Arapaki}}, \bibnamefont{and}
  \bibinfo{author}{\bibfnamefont{P.}~\bibnamefont{Argyrakis}},
  \bibinfo{journal}{Phys. Rev. E} \textbf{\bibinfo{volume}{58}},
  \bibinfo{pages}{2788} (\bibinfo{year}{1998}).

\bibitem[{\citenamefont{Dhar and Stauffer}(1998)}]{dhar1998}
\bibinfo{author}{\bibfnamefont{D.}~\bibnamefont{Dhar}} \bibnamefont{and}
  \bibinfo{author}{\bibfnamefont{D.}~\bibnamefont{Stauffer}},
  \bibinfo{journal}{Int. J. Mod. Phys. C} \textbf{\bibinfo{volume}{09}},
  \bibinfo{pages}{349} (\bibinfo{year}{1998}), \bibinfo{note}{{A summary of
  this article is available at papercore.org, see
  {http://www.papercore.org/Dhar1998}}}.

\bibitem[{\citenamefont{Kirsch}(1999)}]{kirsch1999}
\bibinfo{author}{\bibfnamefont{A.}~\bibnamefont{Kirsch}},
  \bibinfo{journal}{Int. J. Mod. Phys. C} \textbf{\bibinfo{volume}{10}},
  \bibinfo{pages}{753} (\bibinfo{year}{1999}).

\bibitem[{\citenamefont{Zhang et~al.}(2006)\citenamefont{Zhang, W.-Z., Zou, and
  Jin}}]{zhang2006}
\bibinfo{author}{\bibfnamefont{Z.~X.} \bibnamefont{Zhang}},
  \bibinfo{author}{\bibfnamefont{O.}~\bibnamefont{W.-Z.}},
  \bibinfo{author}{\bibfnamefont{X.-W.} \bibnamefont{Zou}}, \bibnamefont{and}
  \bibinfo{author}{\bibfnamefont{Z.-Z.} \bibnamefont{Jin}},
  \bibinfo{journal}{Physica A} \textbf{\bibinfo{volume}{360}},
  \bibinfo{pages}{391 } (\bibinfo{year}{2006}).

\bibitem[{\citenamefont{M. and Dhar}(1983)}]{barma1983}
\bibinfo{author}{\bibfnamefont{B.}~\bibnamefont{M.}} \bibnamefont{and}
  \bibinfo{author}{\bibfnamefont{D.}~\bibnamefont{Dhar}}, \bibinfo{journal}{J.
  Phys. C} \textbf{\bibinfo{volume}{16}}, \bibinfo{pages}{1451}
  (\bibinfo{year}{1983}).

\bibitem[{\citenamefont{Dhar}(1984)}]{dhar1984}
\bibinfo{author}{\bibfnamefont{D.}~\bibnamefont{Dhar}},
  \bibinfo{journal}{Journal of Physics A: Mathematical and General}
  \textbf{\bibinfo{volume}{17}}, \bibinfo{pages}{L257} (\bibinfo{year}{1984}).

\bibitem[{\citenamefont{Pandey}(1984{\natexlab{b}})}]{pandey1984}
\bibinfo{author}{\bibfnamefont{R.~B.} \bibnamefont{Pandey}},
  \bibinfo{journal}{Phys. Rev. B} \textbf{\bibinfo{volume}{30}},
  \bibinfo{pages}{489} (\bibinfo{year}{1984}{\natexlab{b}}).

\bibitem[{\citenamefont{Skarpalezos et~al.}(2013)\citenamefont{Skarpalezos,
  Kittas, Argyrakis, Cohen, and Havlin}}]{Skarpalezos2013}
\bibinfo{author}{\bibfnamefont{L.}~\bibnamefont{Skarpalezos}},
  \bibinfo{author}{\bibfnamefont{A.}~\bibnamefont{Kittas}},
  \bibinfo{author}{\bibfnamefont{P.}~\bibnamefont{Argyrakis}},
  \bibinfo{author}{\bibfnamefont{R.}~\bibnamefont{Cohen}}, \bibnamefont{and}
  \bibinfo{author}{\bibfnamefont{S.}~\bibnamefont{Havlin}},
  \bibinfo{journal}{Phys. Rev. E} \textbf{\bibinfo{volume}{88}},
  \bibinfo{pages}{012817} (\bibinfo{year}{2013}).

\bibitem[{\citenamefont{Caracciolo et~al.}(1992)\citenamefont{Caracciolo,
  Pelissetto, and Sokal}}]{caracciolo1992}
\bibinfo{author}{\bibfnamefont{S.}~\bibnamefont{Caracciolo}},
  \bibinfo{author}{\bibfnamefont{A.}~\bibnamefont{Pelissetto}},
  \bibnamefont{and} \bibinfo{author}{\bibfnamefont{A.~D.} \bibnamefont{Sokal}},
  \bibinfo{journal}{J. Stat. Phys.} \textbf{\bibinfo{volume}{67}},
  \bibinfo{pages}{65} (\bibinfo{year}{1992}).

\bibitem[{\citenamefont{Kremer}(1981)}]{kremer1981}
\bibinfo{author}{\bibfnamefont{K.}~\bibnamefont{Kremer}}, \bibinfo{journal}{Z.
  Phys. B} \textbf{\bibinfo{volume}{45}}, \bibinfo{pages}{149}
  (\bibinfo{year}{1981}), \bibinfo{note}{{A summary of this article is
  available at papercore.org, see {http://www.papercore.org/Kremer1981}}}.

\bibitem[{\citenamefont{Grassberger}(1993)}]{grassberger1993}
\bibinfo{author}{\bibfnamefont{P.}~\bibnamefont{Grassberger}},
  \bibinfo{journal}{J. Phys. A} \textbf{\bibinfo{volume}{26}},
  \bibinfo{pages}{1023} (\bibinfo{year}{1993}).

\bibitem[{\citenamefont{Berretti and Sokal}(1985)}]{beretti1985}
\bibinfo{author}{\bibfnamefont{A.}~\bibnamefont{Berretti}} \bibnamefont{and}
  \bibinfo{author}{\bibfnamefont{A.}~\bibnamefont{Sokal}}, \bibinfo{journal}{J.
  Stat. Phys.} \textbf{\bibinfo{volume}{40}}, \bibinfo{pages}{483}
  (\bibinfo{year}{1985}).

\bibitem[{\citenamefont{Smailer et~al.}(1993)\citenamefont{Smailer, Machta, and
  Redner}}]{smailer1993}
\bibinfo{author}{\bibfnamefont{I.}~\bibnamefont{Smailer}},
  \bibinfo{author}{\bibfnamefont{J.}~\bibnamefont{Machta}}, \bibnamefont{and}
  \bibinfo{author}{\bibfnamefont{S.}~\bibnamefont{Redner}},
  \bibinfo{journal}{Phys. Rev. E} \textbf{\bibinfo{volume}{47}},
  \bibinfo{pages}{262} (\bibinfo{year}{1993}).

\bibitem[{\citenamefont{Majumdar}(1992)}]{majumdar1992}
\bibinfo{author}{\bibfnamefont{S.~N.} \bibnamefont{Majumdar}},
  \bibinfo{journal}{Phys. Rev. Lett.} \textbf{\bibinfo{volume}{68}},
  \bibinfo{pages}{2329} (\bibinfo{year}{1992}).

\bibitem[{\citenamefont{Grassberger}(2009)}]{grassberger2009}
\bibinfo{author}{\bibfnamefont{P.}~\bibnamefont{Grassberger}},
  \bibinfo{journal}{J. Stat. Phys.} \textbf{\bibinfo{volume}{136}},
  \bibinfo{pages}{399} (\bibinfo{year}{2009}).

\bibitem[{\citenamefont{Boyer et~al.}(2004)\citenamefont{Boyer, Miramontes,
  Ramos-Fernandez, Mateos, and Cocho}}]{boyer2004}
\bibinfo{author}{\bibfnamefont{D.}~\bibnamefont{Boyer}},
  \bibinfo{author}{\bibfnamefont{O.}~\bibnamefont{Miramontes}},
  \bibinfo{author}{\bibfnamefont{G.}~\bibnamefont{Ramos-Fernandez}},
  \bibinfo{author}{\bibfnamefont{J.~L.} \bibnamefont{Mateos}},
  \bibnamefont{and} \bibinfo{author}{\bibfnamefont{G.}~\bibnamefont{Cocho}},
  \bibinfo{journal}{Physica A} \textbf{\bibinfo{volume}{342}},
  \bibinfo{pages}{329 } (\bibinfo{year}{2004}).

\bibitem[{ACO()}]{ACO_homepage}
\bibinfo{note}{For an overview of current developments regarding the ACO
  meta-heuristic, see: {http://www.aco-metaheuristic.org}}.

\bibitem[{\citenamefont{Dorigo and Blum}(2005)}]{Dorigo2005_aco}
\bibinfo{author}{\bibfnamefont{M.}~\bibnamefont{Dorigo}} \bibnamefont{and}
  \bibinfo{author}{\bibfnamefont{C.}~\bibnamefont{Blum}},
  \bibinfo{journal}{Theor.\ Comp.\ Sci.} \textbf{\bibinfo{volume}{344}},
  \bibinfo{pages}{243} (\bibinfo{year}{2005}).

\bibitem[{\citenamefont{Dorigo}(1997)}]{Dorigo1997_aco}
\bibinfo{author}{\bibfnamefont{M.}~\bibnamefont{Dorigo}},
  \bibinfo{journal}{IEEE Trans.\ Evol.\ Comp.} \textbf{\bibinfo{volume}{1}},
  \bibinfo{pages}{53} (\bibinfo{year}{1997}).

\bibitem[{\citenamefont{Neumann and Witt}(2010)}]{Neumann2010_aco}
\bibinfo{author}{\bibfnamefont{F.}~\bibnamefont{Neumann}} \bibnamefont{and}
  \bibinfo{author}{\bibfnamefont{C.}~\bibnamefont{Witt}},
  \bibinfo{journal}{Theor.\ Comp.\ Sci.} \textbf{\bibinfo{volume}{411}},
  \bibinfo{pages}{2406} (\bibinfo{year}{2010}).

\bibitem[{\citenamefont{Wilson}(1971)}]{wilson1971}
\bibinfo{author}{\bibfnamefont{K.~G.} \bibnamefont{Wilson}},
  \bibinfo{journal}{Phys. Rev. B} \textbf{\bibinfo{volume}{4}},
  \bibinfo{pages}{3174} (\bibinfo{year}{1971}).

\bibitem[{\citenamefont{Wilson}(1983)}]{wilsonNobelPrizeLecture1983}
\bibinfo{author}{\bibfnamefont{K.~G.} \bibnamefont{Wilson}},
  \bibinfo{journal}{Rev. Mod. Phys.} \textbf{\bibinfo{volume}{55}},
  \bibinfo{pages}{583} (\bibinfo{year}{1983}).

\bibitem[{\citenamefont{Stanley}(1999)}]{stanley1999}
\bibinfo{author}{\bibfnamefont{H.~E.} \bibnamefont{Stanley}},
  \bibinfo{journal}{Rev. Mod. Phys.} \textbf{\bibinfo{volume}{71}},
  \bibinfo{pages}{S358} (\bibinfo{year}{1999}).

\bibitem[{\citenamefont{Melchert and Hartmann}(2011)}]{melchert2011a}
\bibinfo{author}{\bibfnamefont{O.}~\bibnamefont{Melchert}} \bibnamefont{and}
  \bibinfo{author}{\bibfnamefont{A.~K.} \bibnamefont{Hartmann}},
  \bibinfo{journal}{Eur. Phys. J B} \textbf{\bibinfo{volume}{80}},
  \bibinfo{pages}{155} (\bibinfo{year}{2011}).

\bibitem[{\citenamefont{Melchert}(2009)}]{autoScale2009}
\bibinfo{author}{\bibfnamefont{O.}~\bibnamefont{Melchert}},
  \bibinfo{journal}{Preprint: arXiv:0910.5403v1}  (\bibinfo{year}{2009}).

\bibitem[{\citenamefont{Houdayer and Hartmann}(2004)}]{houdayer2004}
\bibinfo{author}{\bibfnamefont{J.}~\bibnamefont{Houdayer}} \bibnamefont{and}
  \bibinfo{author}{\bibfnamefont{A.~K.} \bibnamefont{Hartmann}},
  \bibinfo{journal}{Phys. Rev. {\bf B}} \textbf{\bibinfo{volume}{70}},
  \bibinfo{pages}{014418} (\bibinfo{year}{2004}).

\end{thebibliography}

\end{document}